\documentclass[a4paper,onecolumn,accepted=2024-11-04]{quantumarticle}
\pdfoutput=1
\usepackage[T1]{fontenc}
\usepackage[latin9]{inputenc}
\usepackage{babel}
\usepackage{graphicx,amsmath,amssymb,color}
\usepackage[normalem]{ulem}
\usepackage{amsfonts}
\usepackage[toc,page]{appendix}
\usepackage{hyperref}
\usepackage{latexsym}
\usepackage{amsfonts}
\usepackage{algpseudocode}
\usepackage{amsthm}
\usepackage{mathrsfs}
\usepackage{color,verbatim}
\usepackage{psfrag}
\usepackage[numbers]{natbib}
\usepackage{rotating} 
\usepackage{bbold} 
\usepackage{multirow}
\usepackage{bigstrut} 

\usepackage{subcaption} 

\usepackage[svgnames]{xcolor}
\usepackage{tikz}

\usepackage{algorithm}

\newcommand{\bra}[1]{\langle#1 |}
\newcommand{\ket}[1]{|#1 \rangle}
\newcommand{\bigket}[1]{\Bigl \lvert#1  \Bigr \rangle}
\newcommand{\braket}[2]{\left \langle #1 \middle \vert #2 \right \rangle}

\newcommand{\ketbra}[2]{\vert #1 \rangle \! \langle #2 \vert}

\newcommand{\sandwich}[3]{\left \langle #1 \middle \vert #2 \middle \vert #3 \right\rangle}

\definecolor{myDarkRed}{rgb}{0.5, 0, 0}

\definecolor{myOrange}{rgb}{0.7, 0.3, 0} 
\definecolor{myDarkGreen}{rgb}{0, 0.7, 0} 

\definecolor{myAqua}{rgb}{0, 0.3, 0.3}

\begin{document}

\title{Grover Speedup from Many Forms of the Zeno Effect}
\author{Jesse Berwald$^\dagger$, Nick Chancellor, Raouf Dridi$^\dagger$}
\date{Quantum Computing Inc (QCi), 5 Marine View Plaza Hoboken NJ 07030 USA \newline
$\dagger$ current affiliation: Qamia, ADGM, Abu Dhabi}

\maketitle

\begin{abstract}

It has previously been established that adiabatic quantum computation, operating based on a continuous Zeno effect due to dynamical phases between eigenstates, is able to realise an optimal Grover-like quantum speedup. In other words, is able to solve an unstructured search problem with the same $\sqrt{N}$ scaling as Grover's original algorithm. A natural question is whether other manifestations of the Zeno effect can also support an optimal speedup in a physically realistic model (through direct analogue application rather than indirectly by supporting a universal gateset). In this paper we show that they can support such a speedup, whether due to measurement, decoherence, or even decay of the excited state into a computationally useless state. Our results also suggest a wide variety of methods to realise speedup which do not rely on Zeno behaviour. We group these algorithms into three families to facilitate a structured understanding of how speedups can be obtained: one based on phase kicks, containing adiabatic computation and continuous-time quantum walks; one based on dephasing and measurement; and finally one based on destruction of the amplitude within the excited state, for which we are not aware of any previous results. These results suggest that there may be exciting opportunities for new paradigms of analog quantum computing based on these effects.

\end{abstract}

\tableofcontents

\section{Introduction \label{sec:intro}}

 It is well known that there are many manifestations of the Zeno effect for example, from measurement, from either dissipative or unitary coupling, or from random phase kicks \cite{Misra1977Zeno,Facci2002Zeno,Itano2009ZenoPerspect}. Aside from the well-studied case of adiabatic quantum computation \cite{Facci2002Zeno,Albash2018adiabatic} which arises effectively from a Zeno effect due to continuous phase kicks, it is less clear how useful these different manifestations will be for directly performing analog quantum computation. 

 Many manifestations of the Zeno effect are known to be able to realise universal gatesets (which are in turn able to realise an optimal advantage on unstructured search) in the setting of digital gate-model quantum computation \cite{Huang2008zenouniversal, Beige2000dissdecohere}. Our work instead focuses on using these effects in a direct analog way, akin to the role of dynamic phases within adiabatic quantum computation. 

Beyond just being a theoretical construct, demonstrations have been made of the ability of Zeno effects to preserve subspaces.  A continuous ``partial measurement'' on a transmon system \cite{Hatridge2013Transmon} was shown to realise a Zeno effect experimentally. Beyond this, quantum Zeno dynamics; dynamics within a non-trivial subsystem isolated using the Zeno effect was demonstrated experimentally within an atomic system \cite{Signoles2014confined}, with similar effects soon demonstrated for photons \cite{Bretheau2015confined}. Relatedly, a kind of Zeno effect was demonstrated to be capable of reaching high success probability in an interaction free measurement protocol on transmon systems \cite{Dogra2022interactionfree}. For a review of Zeno dynamics, see \cite{Facci2008Zeno}. 

Definitions of the Zeno effect and Zeno dynamics may vary slightly. For the purpose of the present paper by Zeno effect, we mean any interaction which prevents a quantum system from entering a state by interfering with buildup of coherences, which can be expressed mathematically as the off-diagonal elements of a density matrix. Likewise, by Zeno dynamics, we mean the (possibly trivial) dynamics of a system which is confined by Zeno effects.
 
 If one is to consider analog computing devices based on novel mechanisms, verifying that a mechanism has potential for a speedup is a key step to validate it as a useful tool for quantum computing. In this manuscript, we focus on a speedup on unstructured search. The advantage of this setting, is that the scaling of the best known possible algorithm is known, so an advantage can be proven. This contrasts with combinatorial optimisation problems, where the best classical scaling is not known. Of particular relevance to this work is the recently proposed entropy computing paradigm of photonic computing \cite{nguyen2024entropycomputing}, which aims to build analog devices making use of a number of effects, including Zeno blockade \cite{huang2012antibunched,McCusker_2013, Huang2010}. Entropy computing is an optical paradigm which can support a fully quantum implementation and has demonstrated promising results in early experiments which make use of classical electronics for feedback \cite{nguyen2024entropycomputing}. 
 
 Conceptually, entropy computing is similar to the coherent Ising machine paradigm \cite{nguyen2024entropycomputing,Yamamoto2017CoherentIsing} but differs in two key aspects. Firstly, encoding is performed directly in amplitudes, as opposed to quadrature, and operation in the few-photon regime is possible. This potentially alleviates some issues with uncontrolled amplitude fluctuations disrupting an accurate quadrature encoding \cite{Yamamoto2017CoherentIsing}. 

 The second key way in which entropy computing differs from the coherent Ising paradigm is the fact that both loss and gain are controlled in the entropy computing paradigm; while only controllable gain is used in the coherent Ising setting. This controllable loss is where the Zeno blockade features in the entropy computing paradigm. By using precisely controlled fabrication techniques, non-linear optics can be used to engineer extremely high photon loss rates, which can lead to Zeno-mediated interactions even at the single photon level. Crucially, in a certain operating regime, a Zeno blockade system can exhibit the underlying effects for one of the computational models we discuss later. Because of the background required, we reserve the details for section \ref{sec:discussion}.
 
 It is worth briefly pointing out that many implementations of coherent Ising machines are based on classical feedback which destroys meaningful quantum superposition \cite{Honjo2021CIM} in this case one would expect the classical scaling for unstructured search. In fact, simulations can sometimes out-perform the physical machines \cite{Tiunov2019CIMinspired}. Similarly the currently implemented entropy computers use such feedback \cite{nguyen2024entropycomputing}. Both paradigms however can incorporate quantum search if implemented in an all-optical way \cite{nguyen2024entropycomputing,Yamamoto2017CoherentIsing}.

Additionally, Zeno effects have been known to have various applications in quantum computing, for example in this proposal for constructing a universal quantum computer \cite{PhysRevA.77.062332}. Therein the authors demonstrate that interaction-free imaging can be extended to the few-atom level, enabling the realization of asymptotically on-demand interaction- and measurement-free quantum logic gates, which are robust against decoherence and detector inefficiency. Interaction-free Zeno gates were experimentally demonstrated in \cite{Blumenthal2022Zeno} building on methods proposed in \cite{Burgarth2014Zeno}. Holonomic quantum gates are often designed and implemented based on adiabatic effects to preserve a subspace \cite{Zanardi1999holonomic,Chancellor2013holonomic,Zhang21holonomic}. Strong dissipation can also be used to maintain a system within a decoherence free subspace \cite{Beige2000dissdecohere}. Furthermore, \cite{Huang2008zenouniversal} demonstrated that dissipation based Zeno gates can support universal quantum computation. More recently, \cite{mi2023stable} showed that engineered dissipative reservoirs can guide many-body quantum systems toward correlated steady states, facilitating scalable entangled many-body states and the exploration of nonequilibrium quantum phenomena. Zeno effects also have applications in error correction~\cite{Paz-Silva2012Zeno_error}. In \cite{PyshkinZenoGrover2022} it was even demonstrated that a non-unitary Zeno effect implemented through measurement can be used to obtain a Grover-type advantage, a specific example of the more general result we show here. Zeno effects have also been found to be useful in implementing constraints in a gate-model setting \cite{Herman2023constrainedZeno,dizaji2024zenosubspaces,liu2024comparisonconstrainencodingmethods}. Universal bounds for error rates have been derived \cite{Burgarth2022oneboundtorulethem}.

More broadly, even beyond Zeno effects, open quantum system effects can have a positive and useful effect on a system. For example, improving aspects such as uniformity in quantum walks \cite{kendon2007decoherereview}. Beneficial effects from open-system interactions have also been directly used in technology. For example, the reverse annealing feature as implemented in D-Wave systems only works because of finite temperature dissipation \cite{chancellor17b,chancellor21a}. For another example of the beneficial effect of open system dynamics in that setting, see \cite{dickson13a}. Consistent with these trends. Our work suggests that there will be many optimal protocols which rely on open system effects but are not directly manifestations of Zeno dynamics.

While real optimisation problems of interest have structure which should not be ignored, performance on unstructured search can give an indication whether it is hypothetically possible to use quantum superposition to gain an advantage. In this work we focus on whether different computational mechanisms, corresponding to different versions of Zeno effects (both continuous and discrete) can provide a quantum advantage on unstructured search. We find that for direct analog quantum computation based on all mechanisms we examine, projective measurements, phase kicks, and a form of measurement inspired by dissipation which we call destructive measurement a quantum speedup is possible. Furthermore, all models can reach the theoretically optimal $\sqrt{N}$ speedup, matching the scaling of the original algorithm by Grover \cite{Grover1996search,Grover1997Search}. While the continuum limit of a phase kicks correspond to the adiabatic algorithm, which has already been shown to attain an optimal speedup on unstructured search \cite{Roland2002}, many of the others have not been explored. This observation highlights opportunities for novel methods of performing quantum optimisation. 

Unstructured search provides a natural starting place for understanding algorithms, since it operates in a relatively simple two-state subspace and has an understandable advantage. In fact, unstructured search performance provided key evidence for the utility of continuous time quantum walk \cite{Childs2004spatial} and adiabatic quantum computing \cite{Roland2002}. Although other evidence also played an important role, such as \cite{childs2003exponential,farhi98qwdecision} where an exponential advantage in graph traversal and some decision trees were demonstrated for continuous-time quantum walks.

Additionally, unstructured search is where the exploration between quantum annealing and continuous-time quantum walks began \cite{Morley19a}. Continuous time quantum walks now provide an important tool for understanding quantum annealing far from the adiabatic limit. Based on this ability to interpolate in a way that maintained an advantage, there grew the idea that continuous-time quantum walks could be useful for optimisation problems with structure \cite{Callison2019QW} (see also \cite{Hastings19a} and \cite{Chancellor2020perspectiveduality}). From this point it became clear that energetic arguments of the type often used in continuous-time quantum walk could also be applied to multi-stage quantum walks, and this could be taken to a continuum limit which represented rapid quenches \cite{Callison21a}. Continuous time quantum walks have since provided a key reference point for understanding diabatic quantum annealing \cite{crosson2020prospects,Banks2024continuoustime,Banks2024rapidquantum,Schulz2024guided}. In this work we hope to provide a similar starting place for theoretically understanding the more general Zeno-based hardware being explored in the entropy computing paradigm \cite{nguyen2024entropycomputing}. Extensions of the understanding presented here have already begun in \cite{berwald2023zeno}. 

We structure the presentation of our work as follows. In section \ref{sec:back_strat} we review the mathematical background and summarise the strategy we will use for deriving our results. Details of unstructured search Hamiltonians and the ``single avoided crossing'' model which we use can be found in appendices \ref{appendix:unstruct_search} and \ref{appendix:avoided_cross}. In all of these examples, the systems are subject to various quantum channels. In section \ref{sec:gap_invaraint} we lay out the key aspects of our strategy, the conversion of the single avoided crossing model to one which is gap invariant by scaling total runtime with the gap. The next three section discuss their individual families of algorithms. In section \ref{sec:phase_rot} we discuss a family of algorithms based on phase rotation, where many protocols are already known, but we add some more. Section \ref{sec:dec_family} discusses a family of algorithms based on decoherence-like quantum channels which is not well explored but does include search by measurement. Section \ref{sec:dest_family} discusses a family of algorithms based on destruction of the excited state, something which to our knowledge has not yet been discussed in the literature. We synthesise these results into an overall picture of the three families in section \ref{sec:discussion}.

\section{Mathematical Strategy\label{sec:back_strat}}

\subsection{Model background}

We first define methods which are able to solve the unstructured search problem using a single quantum channel acting nontrivially on the basis states in the space defined by the two vectors $\{\ket{\tilde{\omega}},\ket{m}\}$, where $\ket{m}$ is the marked state and $\ket{\tilde{\omega}}$ is a uniform superposition of all other classical basis states defined as
\begin{equation}
    \ket{\tilde{\omega}}=\frac{1}{\sqrt{1-\left|\braket{\omega}{m} \right|^2}}\left( \ket{\omega}-\braket{m}{\omega}\ket{m} \right),
\end{equation}
where $\ket{\omega}$ is an equal positive superposition of all computational basis states. 

The quantum channel used to solve the problem could for example be applying phases, making a projective measurement, applying decoherence, or allowing the first excited state to dissipate into a computationally useless state we call $\ket{d}$. A crucial step here is to argue over what timescales this quantum channel is physically achievable, as was done in \cite{Childs2002measurement} for measurements in this basis. Given that building on the prior work in this paper is crucial to our arguments, we review the key elements of the model from \cite{Childs2002measurement} in appendix \ref{appendix:comp_by_measure}.  Making arguments about the allowed timescales of these to implement the necessary quantum channels is a crucial step, if not done carefully it could lead to incorrect conclusions. For example, if we assume that dephasing or dissipation can be accomplished in constant time, independent of the size of the system, then this would lead to a system where unstructured search of any size can be accomplished in the same amount of time. Such a result can only come from an unrealistic description of a physical system: any physical quantum system can be efficiently simulated by a universal quantum computer, and it has been proven that such computers cannot exceed the $\sqrt{N}$ scaling of Grover's algorithm \cite{Bennett1997GroverBest}. Such a system would have to be ``unphysical'' or it would contradict a known result. On the other hand, it is not \emph{a priori} given that every quantum channel should result in any speedup over the classical methods. A major contribution of the current work is to show that for the models we examine, physically realistic versions are able to achieve (but not exceed) $\sqrt{N}$ scaling.

To build a Zeno effect based on discrete measurements in the present model, we must consider a continuously changing basis, which can be defined by the ground and first excited states of an unstructured search Hamiltonian. We then argue that a large but fixed (with respect to the problem size) number of repeated implementations of a quantum channel occurring within the region of parameter space where the marked state and uniform superposition both have significant overlap with the ground state will maintain the original $\sqrt{N}$ scaling while faithfully describing a Zeno effect. We then further argue the existence of a continuum limit which will also retain the scaling and describe a time dependent quantum channel which implements a Zeno effect.

This work builds on a model of unstructured search known as the single avoided crossing model. The model we use was formally introduced in \cite{Morley19a}, but the underlying concept has been implicit in many other works \cite{Roland2002,Farhi1998QuantumWalk,Childs2004spatial,Chakraborty2016randGraphs}. This model is relevant to a wide variety of cases of unstructured search, including all computationally relevant qudit models, and for both transverse field (aka hypercube) and complete graph drivers. A full discussion of the relevant search Hamiltonians appears in appendix \ref{appendix:unstruct_search}.

The model from \cite{Morley19a} is defined in terms of a two level system,
\begin{equation}
    H_{\mathrm{ac}}=\frac{g_{\mathrm{min}}}{2}\left[f(\tau)Z-X\right], \label{eq:H_ac}
\end{equation}
where $Z=\left( \begin{array}{cc}1 & 0 \\ 0 & -1 \end{array} \right) $, $X=\left( \begin{array}{cc}0 & 1 \\ 1 & 0 \end{array} \right) $, $0\le \tau \le 1$ is a dimensionless time parameter, such that $t=\tau \,t_{\mathrm{scale}} g^{-1}_\mathrm{min}$ where $t_{\mathrm{scale}} g^{-1}_{\mathrm{min}}$ defines an overall time scale and $g_\mathrm{min}$ is the minimum gap. The function $f(\tau)$ is a control function which defines how the Hamiltonian changes with time with boundary conditions that $f(0)= \infty$ and $f(1)= -\infty$ The prefactor $g_{\mathrm{min}}$ is the minimum spectral gap. This Hamiltonian is mapped such that $\ket{\tilde{\omega}}\rightarrow \ket{1}$ and $\ket{m}\rightarrow \ket{0}$. Within the single avoided crossing model, an optimal annealing schedule is defined as
\begin{equation}
f(\tau)=\cot(\pi \tau). \label{eq:opt_sched}
\end{equation}
Derivations of this quantity, plus a full exact solution  of the single avoided crossing model, appear in appendix \ref{appendix:avoided_cross}. A key quantity here is the energy gap between the ground and first excited state,
\begin{equation}
    g(\tau)=g_\mathrm{min}\sqrt{f^2(\tau)+1} \label{eq:gap_ac}.
\end{equation}
We note that, taken naively, this model appears to include terms with an infinite energy scale as $\tau \rightarrow 0$ and $\tau \rightarrow 1$. Such divergences, while they appear alarming, are actually appropriate given that $f(\tau)$ is rescaled by $g_\mathrm{min}$. The realistic condition is that away from the avoided crossing the total energy scale $f(\tau)g_\mathrm{min}= O(1)$. If we applied a version which includes a cutoff 
\begin{equation}
\bar{f}(\tau)=\begin{cases} f(\tau) & |f(\tau)|<g^{-1}_\mathrm{min} \\ \mathrm{sgn}(f(\tau))\,g^{-1}_\mathrm{min} & \mathrm{otherwise} \end{cases}
\end{equation}
the results obtained would be identical. Firstly, we can see that the proportion of $\tau$ values where this cutoff is relevant would vanish very quickly it would only be relevant when $\cot^{-1}(g^{-1}_\mathrm{min})/\pi>\tau$ or $\cot^{-1}(-g^{-1}_\mathrm{min})/\pi<\tau$. Based on the Laurante expansion of the cotangent function, these correspond approximately to $\tau<g_\mathrm{min}$ and $\tau>1-g_\mathrm{min}$. Since $g_\mathrm{min}$ scales as $N^{-\frac{1}{2}}$, the range where this cutoff is relevant quickly vanishes. Furthermore, beyond this cutoff the eigenstates of the system are to a very good approximation just the marked state and equal superposition. Performing a perturbative expansion, we find that in fact where $f(\tau)\approx g^{-1}_\mathrm{min}$, the eigenstates of the system will be $\{\ket{m}+O(g^{2}_\mathrm{min})\ket{\tilde{\omega}},\ket{\tilde{\omega}}+O(g^{2}_\mathrm{min})\ket{m}\}=\{\ket{m}+O(N^{-1})\ket{\tilde{\omega}},\ket{\tilde{\omega}}+O(N^{-1})\ket{m}\}$. This suggests further that the error in the populations from using $f(\tau)$ instead of the more realistic $\bar{f}(\tau)$ will be of $O(N^{-1})$, which is no larger than the level of error we have already implicitly accepted from ignoring the possibility of randomly guessing the correct solution.  For this reason, we can safely proceed with a model based on $f(\tau)$ despite the diverging energy values as we can be assured that the error caused by this approach could \emph{at most} be as much as neglecting the possibility of random guessing. The relative phases will be quite different between the models, but the population is what is relevant for the computation.

 \subsection{Invariant Strategy\label{sec:gap_invaraint}}

\begin{figure}
    \centering
    \includegraphics[width=9 cm]{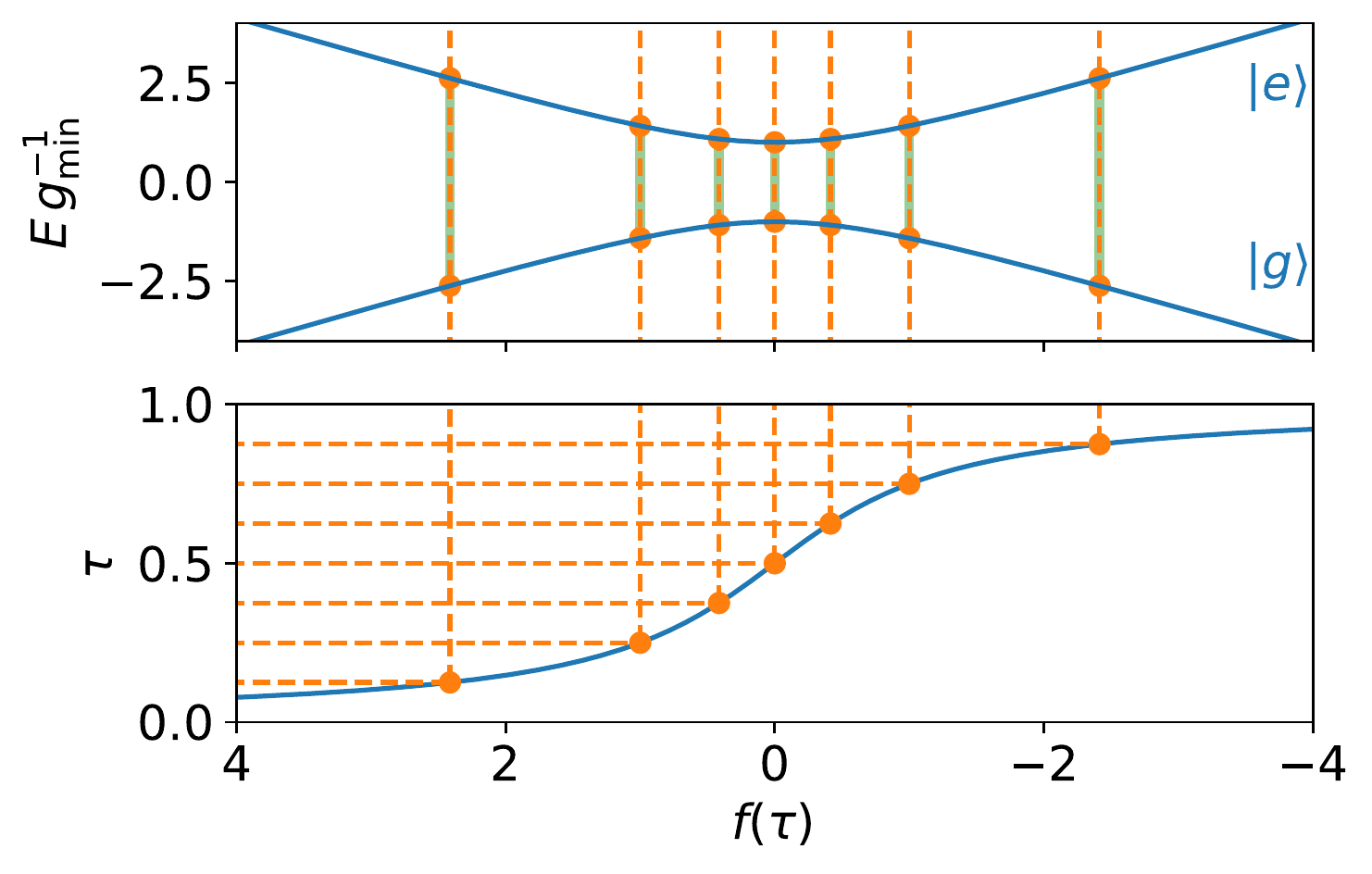}
    \caption{Illustration of how an optimal annealing schedule can be used to decide parameters to perform discrete operations (measurements, for example). In this example, seven evenly spaced values of $\tau$ (placed at $1/8$ increments in a way which excludes $0$ and $1$) are given to an optimal schedule taking the form of equation \ref{eq:opt_sched}. These values of $f(\tau)$ then determine the values, as the top figure shows this leads to concentration around the smallest gap (emphasised with faint green line).}
    \label{fig:sched_illustration}
\end{figure}

 The single avoided crossing model can generally be invoked for any operation which occurs on a timescale proportional to the inverse of the instantaneous gap $g^{-1}(\tau)$. These certainly include adiabatic and quantum walk protocols, but we will demonstrate that it also includes many other operations. Once time scaling with $g^{-1}(\tau)$ is established, then a family of operations which are invariant with the scaling of $g_\mathrm{min}$ can be established, these can either consist of a series of discrete operations, or a continuum quantum channel. Recall that we are interested in the dynamics induced in a two state subspace where the relevant energy scales can be defined by $H_{\mathrm{ac}}$, as discussed in more detail later. Since the purpose of this paper is optimal unstructured search using Zeno dynamics, we focus on the subset of discrete and continuous protocols based on quantum channels which map to Zeno effects (it is however worth noting that our work illuminates many non-Zeno protocols which will also yield an optimal speedup). The final step is to verify that a given protocol yields a success probability which is non-zero,\footnote{within the single avoided crossing model, which does not include the exponentially vanishing probability of randomly guessing the correct solution} which can be argued either analytically or numerically. Generally any non-trivial dynamics within the relevant subspace should accomplish this goal. The process for constructing a protocol which give an optimal speedup is as follows, operating the protocol with a total runtime $T\propto g^{-1}_\mathrm{min}\propto \sqrt{N}$:
\begin{enumerate}
    \item Identify an implementation of a quantum channel where the relevant timescale is proportional to $g^{-1}(\tau)$
    \item Develop a protocol consisting of a series of implementations of this channel which remains mathematically identical as long as $T g_\mathrm{min}$ is held fixed. In other words $H_{\mathrm{ac}}(\tau) T$ is scale invariant with $g_\mathrm{min}$ for all values of unitless time $\tau$. 
    \begin{itemize}
        \item In this paper we focus on protocols which exhibit Zeno effects, but there will also be many non ``Zeno-like'' protocols which exhibit an optimal speedup
    \end{itemize}
    \item Verify that the protocol has a non-zero probability of finding $\ket{m}$ starting from $\ket{\tilde{\omega}}$ within the single avoided crossing model.
\end{enumerate}
We base our protocols on the known optimal adiabatic schedule given in equation \ref{eq:opt_sched} because it provides a method to guarantee either that the protocol slows down appropriately near the crossing in the case of continuous algorithms, or that operations concentrate appropriately around the crossing in the case of protocols based on discrete operations. Note that since $H_{\mathrm{ac}}\propto g_\mathrm{min}$ is fixed we have implicitly set $T\propto g_\mathrm{min}^{-1}$ (corresponding to the optimal Grover runtime), and hence do not need to directly reference it again. For discrete protocols we take evenly spaced values of $\tau$, an example of how these points are chosen is presented visually in figure \ref{fig:sched_illustration}.

While all examples here arise from a quantum channel which can achieve $O(1)$ success probability on a timescale proportional to $g^{-1}(\tau)$, it is an interesting hypothetical to consider what other scaling with $g(\tau)$ could mean. Any scaling which is slower than $g^{-1}(\tau)$ would correspond to a speedup better than the known optimal (Grover) speedup, and therefore could only come about from a quantum channel which is somehow unphysical (we give an example later where an unphysical assumption about measurement accuracy appears to show such a speedup). Scaling which is faster than $g^{-1}(\tau)$, but slower than $g^{-2}(\tau)$, would correspond to a non-optimal speedup. A quantum channel where the runtime scales as $g^{-2}(\tau)$ would result in an overall algorithmic scaling of $N$, which is the best classical scaling. Finally, $g^{-2}(\tau)$ or worse would correspond to no speedup over classical scaling.

We examine the implementation of three types of quantum channels and obtain three families of algorithms, the first family is algorithms where the driving mechanism is \textbf{phase rotation}. This includes the familiar and well understood algorithms of adiabatic quantum computing and continuous time quantum walks. While this family is well understood with already proven speedups, we introduce a few minor additions, multi-stage quantum walks and discrete operations we call ``phase flipping''. 

The second family of algorithms we study is the family of algorithms where the driving mechanism is \textbf{dephasing}. Since projective measurement is a kind of dephasing, this includes the known result of \cite{Childs2002measurement} but is otherwise not well explored.

The third and final family we study is one based on \textbf{destruction} or dissipation of the excited state into a computationally useless state. This is effectively either continuous or discrete implementation of a channel which moves amplitude from an excited state $\ket{e}$ to a computationally useless ``destroyed'' state $\ket{d}$
\begin{equation}
    \rho \rightarrow \ketbra{d}{e}\rho \ketbra{e}{d}+\ketbra{g}{g}\rho \ketbra{g}{g}+\ketbra{d}{d}\rho \ketbra{d}{d}.
\end{equation}

As far as we are aware, this is a class of algorithms which have not previously been explored in the literature. It is however reminiscent of qubit leakage models which are commonly considered in superconducting qubits \cite{Gambetta2011leakage,Chen2016leakage} (although relevant to any model which builds qubits from weakly anharmonic oscillators). These models however, act on single physical qubits, not collective subspaces. We use the term destruction rather than dissipation to refer to this family of algorithms, to avoid confusion with computation driven by dissipation from the excited state into the ground state, an important topic, but not one which we address in the present work because it cannot be considered a source of a Zeno effect.

\section{Phase Rotation Family of Algorithms \label{sec:phase_rot}}

\subsection{Applying our Strategy to Quantum Walk and Adiabatic settings}

While a quadratic speedup was already demonstrated for adiabatic quantum computation in \cite{Roland2002}, it is a useful demonstration of our theoretical methods to re-derive this known result using the tools we have developed. This method will also illustrate that multi-stage quantum walks as discussed in \cite{Callison21a} can achieve an optimal speedup. 

To start with, we consider a continuous time quantum walk performed at an arbitrary point  $f(\tau)=\gamma$. The state of the system as a function of time takes the form
\begin{equation}
    \ket{\psi(t)}=\exp\left(\frac{-i g_\mathrm{min}t}{2}\left[\gamma Z -X\right]\right)\ket{\psi(t=0)}.
\end{equation}
By definition, the timescale of this evolution is the inverse gap between the ground and first excited state, which scales with $g^{-1}_\mathrm{min}$. 

From here we can argue that the invariance property we have discussed in section \ref{sec:gap_invaraint}, holds for $t=\tau g^{-1}_\mathrm{min}t_\mathrm{scale}$ leaving a version of the evolution which is invariant with respect to $g_\mathrm{min}$ and therefore $N$,
\begin{equation}
    \ket{\psi(\tau)}=\exp\left(\frac{-i \tau t_\mathrm{scale}}{2}\left[\gamma Z -X\right]\right)\ket{\psi(t=0)}. \label{eq:quantum_walk}
\end{equation}
It is further clear that except for very special values of $\gamma$ and $t$, the final overlap with $\ket{m}$ is non-zero; however, making sure this value does not vanish as $N$ is scaled is key to demonstrate a speedup. A quantum walk at exactly the avoided crossing or within a fixed finite multiple of $g_\mathrm{min}$ of the avoided crossing will always attain an optimal quantum speedup, but this requires exponentially increasing precision. 

\begin{figure}
    \centering
    \includegraphics[width=9 cm]{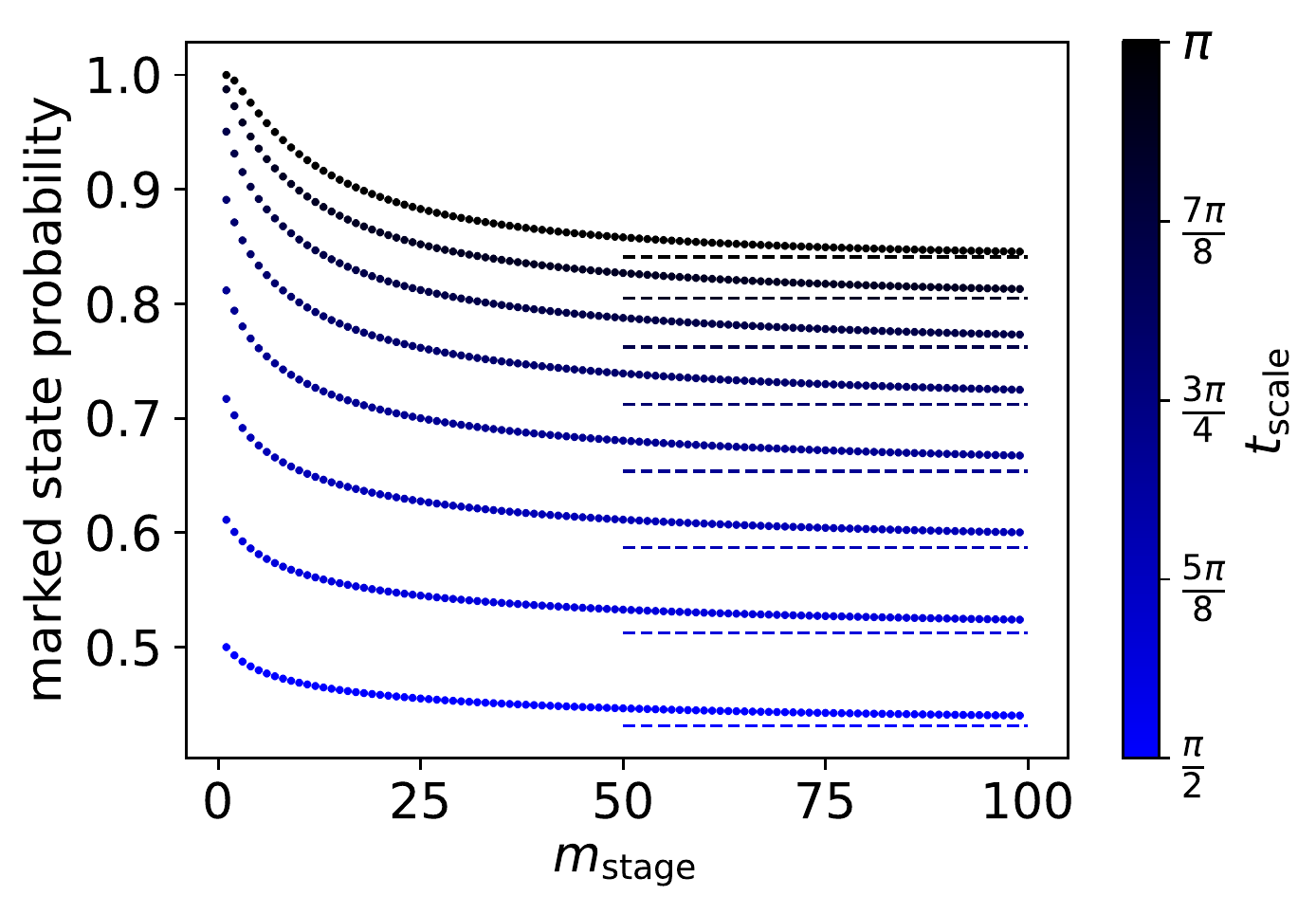}
    \caption{Probability of finding the ground state for a multi-stage quantum walk against $m_\mathrm{stage}$. Points and dotted lines are colour (or grayscale) coded based on different values of $t_\mathrm{scale}$. Dashed lines represent the adiabatic limit which each set of points is approaching, computed by setting $m_\mathrm{stage}=10,000$. The fact that the dashed lines begin at $m_\mathrm{stage}=50$ is an aesthetic choice and has no underlying mathematical significance.}
    \label{fig:msqw_mult_angle}
\end{figure}

It can further be observed that if a finite number of stages of quantum walks were performed \cite{Callison21a} (each with runtimes  $t_{\mathrm{scale},j}$ and using the parameter $\gamma_j=f(\tau_j)=f(j/(m_\mathrm{stage}+1))$, $\mathcal{T}$ is included to remind us of the time ordered nature of the product of non-commuting terms),
\begin{equation}
\ket{\psi(t_\mathrm{scale})}=\mathcal{T}\prod^{m_\mathrm{stage}}_{j=1}\exp\left(\frac{-i t_{\mathrm{scale},j}}{2}\left[\gamma_j Z -X\right]\right)\ket{\psi(t=0)}, \label{eq:trotter_evolution}
\end{equation}
an optimal speedup would still occur, outside of a pathological set of cases of measure zero where the final probability of $\ket{m}$ is exactly zero. Since the subject of this paper is Zeno effects, we focus on subsets of quantum walks which have a Zeno limit. In other words, sequences where scaling up $m_\mathrm{stage}$ will approach a probability of $1$ of being found in the $\ket{m}$ state at the end of the protocol (as both the number of states and the total sum of scaled runtimes becomes large). Even under these restricted circumstances, many such protocols exist. A natural choice is to base the spacing of the quantum walks on the optimal annealing schedule as depicted in figure \ref{fig:sched_illustration} by setting,
\begin{align}
    \gamma_j=\cot(\pi \tau_j)=\cot(\pi \frac{j}{m_\mathrm{stage}+1}), \label{eq:gamma_multistage} \\
    t_{\mathrm{scale},j}=\frac{t_\mathrm{scale}}{m_\mathrm{stage}}\label{eq:t_multistage}.
\end{align}
Probability of finding the marked state for different values of $t_\mathrm{scale}$ and $m_\mathrm{stage}$ are plotted in figure \ref{fig:msqw_mult_angle}. These show that for a fixed runtime where $t_\mathrm{scale}\le \pi$ and no noise, a quantum walk is superior to adiabatic in terms of absolute performance, consistent with the findings of \cite{Morley19a}. We further see that the decrease is monotonic.

\begin{figure}
    \centering
    \includegraphics[width=10 cm]{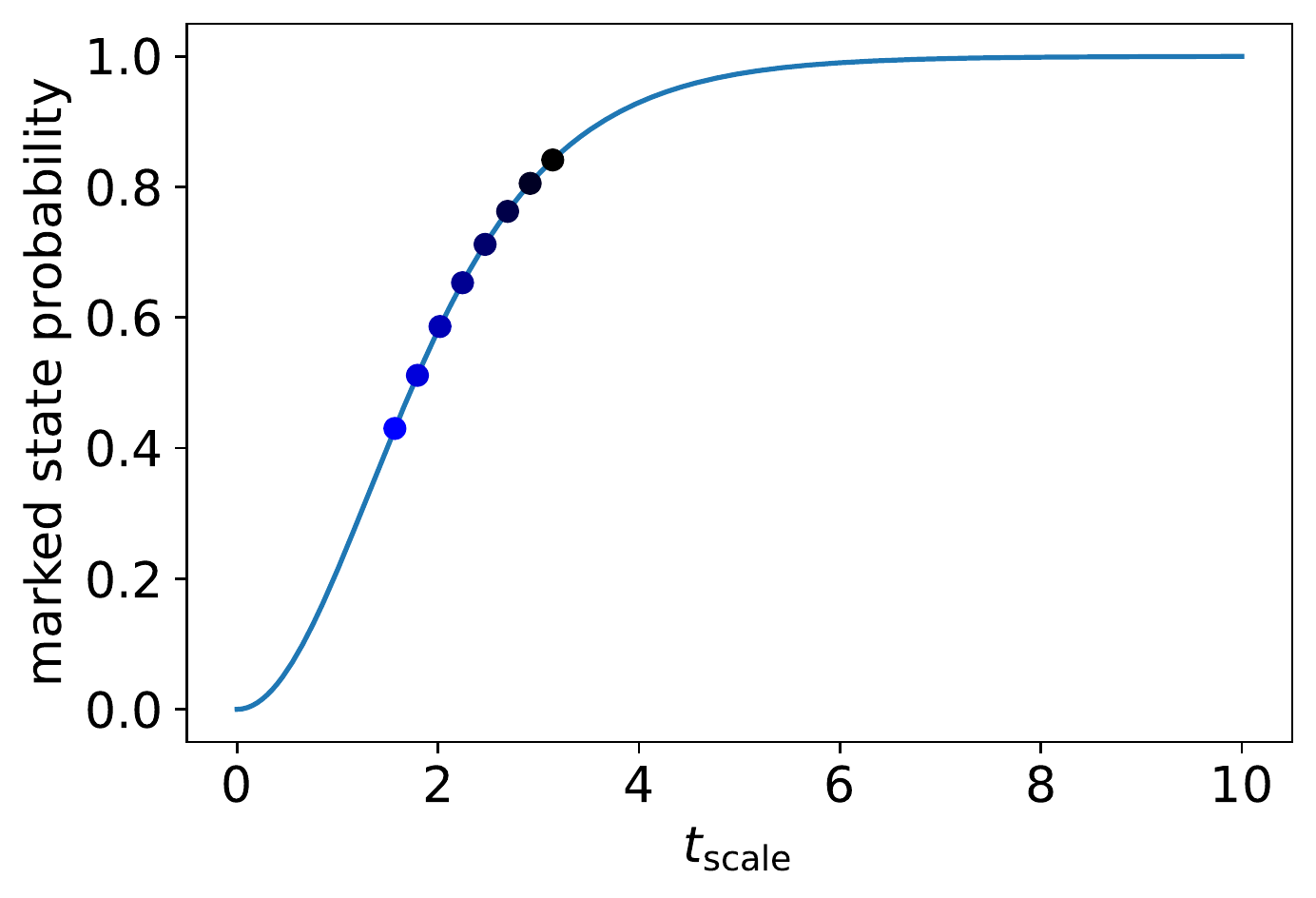}
    \caption{Marked state probability of an adiabatic protocol, which can be defined mathematically as the limit of a multi-stage quantum walk when $m_\mathrm{stage}\rightarrow \infty$ versus $t_\mathrm{scale}$. Approximated numerically by setting $m_\mathrm{stage}=10,000$. Points represent the times used in figure \ref{fig:msqw_mult_angle} color coded to match that plot.}
    \label{fig:adiabat_success_t}
\end{figure}

A feature of the way in which equations \ref{eq:gamma_multistage} and \ref{eq:t_multistage} are defined is that the total scaled runtime $\sum^{m_{\mathrm{stage}}}_jt_{\mathrm{scale},j}=t_\mathrm{scale}$ does not depend on $m_{\mathrm{stage}}$ and therefore the runtime remains finite as $m_{\mathrm{stage}}\rightarrow \infty$. In addition to the results shown in the plots, which by themselves demonstrate an optimal advantage by virtue of being non-zero, it is straightforward to argue that an optimal advantage will be maintained in this limit. This can be done by noting that as the number of stages approaches infinity, a fixed fraction of the time spent when applying such a protocol will be between $\gamma_\mathrm{range}$ and $-\gamma_\mathrm{range}$. Specifically, this fraction will be 
\begin{equation}
    \frac{1}{\pi}\left[\cot^{-1}(\gamma_{\mathrm{range}})-\cot^{-1}(-\gamma_{\mathrm{range}})\right]=\frac{2}{\pi}\arctan(\gamma_\mathrm{range}).
\end{equation}
Since within this range, the eigenstates of the Hamiltonian have an overlap with both $\ket{\omega}$ and $\ket{m}$ this implies that the probability to be found in $\ket{m}$ will be non-zero starting from $\ket{\omega}$ with the possible exception of special runtimes where perfect destructive interference could occur. This proves that with the possible exceptions of special cases, these protocols will give optimal Grover speedup as well. Near the adiabatic limit, the system will approximately follow the instantaneous ground state. The success probability versus $t_\mathrm{scale}$ in the adiabatic limit, in other words, when $m_\mathrm{stage}\rightarrow \infty$ is shown in figure \ref{fig:adiabat_success_t}. 

The structure of the argument makes it clear that within this rescaled picture, a wide variety of schedules will yield an optimal speedup. In fact, such a schedule need only have $f(\tau)$ finite for a finite fraction of $\tau$ values, and to avoid special cases where exact destructive interference could occur. This is a more general restatement of the result of \cite{Morley19a}. The crucial ingredient for optimal speedup is for the schedule to scale in the correct way around the avoided crossing as the problem grows. Other details are not particularly important.

Finally, to unify the discussion of the quantum-walk/adiabatic family of algorithms with the other two cases we study, we generate a single discrete operation which acts as a quantum walk timed to obtain the maximum difference from the initial state. This ``phase flipping'' operation can be thought of as playing an analogous role which projective measurement plays to decoherence. Applying a quantum walk for a time proportional to $\phi/g(\tau)$ will always give the same relative phase (what we call the ``rotation angle''), where $\ket{e}$ acquires a factor of $\exp(i\phi)$. This can be thought of as applying $\tilde{Z}=\ketbra{g}{g}+\exp(i\phi)\ketbra{e}{e}$. 

\begin{figure}
    \centering
    \includegraphics[width=10 cm]{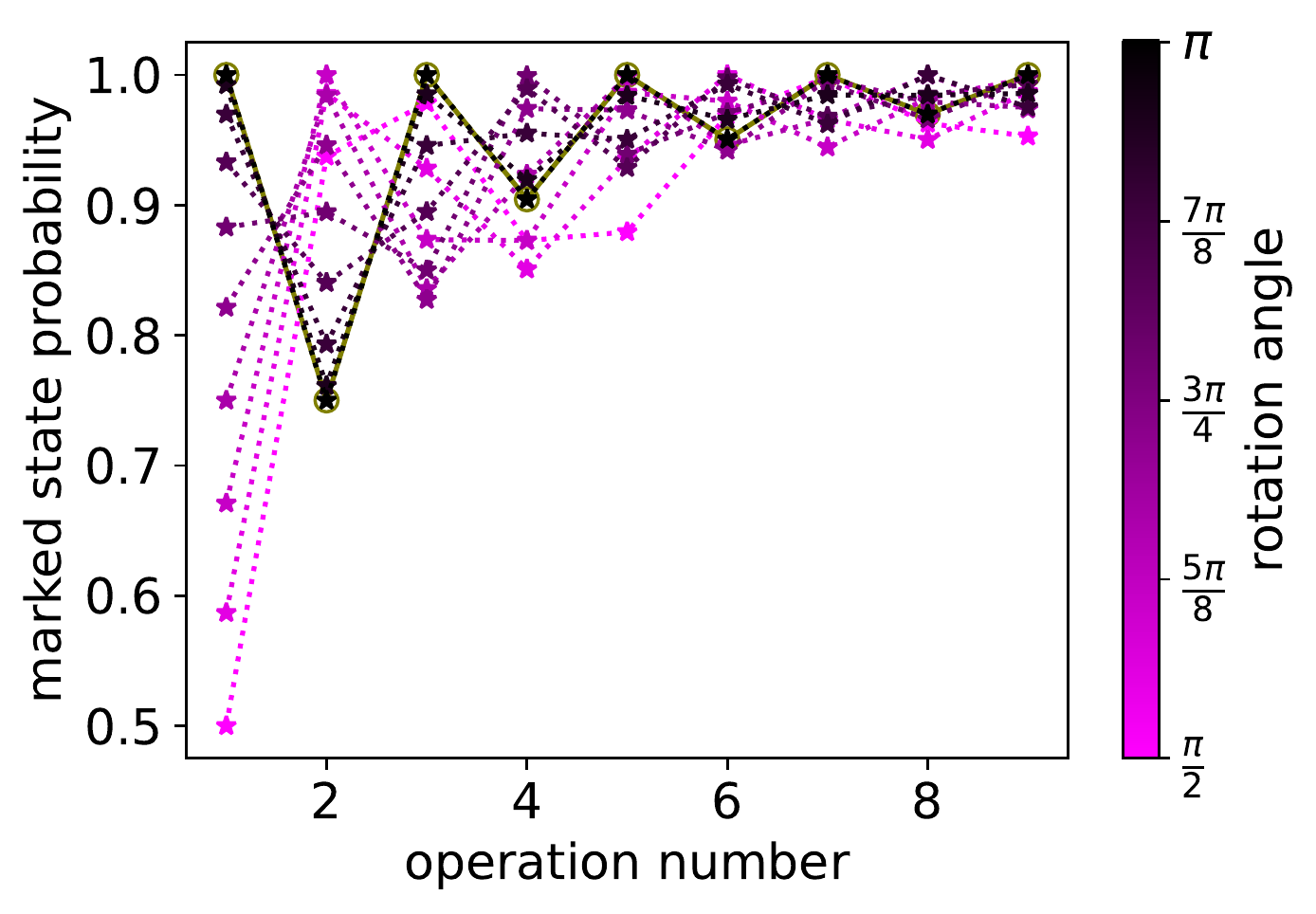}
    \caption{Marked state probability illustrating approach of Zeno limit by performing multiple phase flipping operations (x-axis), each with a fixed rotation angle (colour coding) applied at points defined by an optimal adiabatic schedule. Dotted lines are a guide to the eye. Gold circles indicate exact phase inversion, which as the figure shows is equivalent to a rotation angle of $\pi$.}
    \label{fig:walk_flip_approach}
\end{figure}

We focus on phase flipping, where $\phi=\pi$. A special case is where this operation is performed on state $\ket{\tilde{\omega}}$ at the point where $f=0$. This will result in the state $\ket{m}$ with unit probability. While discrete operations of this nature seem somewhat artificial in the quantum-walk setting, the other two families we discuss will be defined starting from discrete measurement operations. Defining phase flipping, and more generally phase rotation as the primitive element of this family of algorithms is natural in the broader setting. As figure \ref{fig:walk_flip_approach}, shows, multiple phase rotations tend to illustrate a Zeno effect (as seen by the approach to unit marked state probability), but the dependence of the success probability on the number of phase rotation operations is far from monotonic. We also derive the Zeno effect mathematically in appendix \ref{appendix:alg_op}.

\section{Decoherence Family of Algorithms\label{sec:dec_family}}
\subsection{Zeno Effect Induced by Projective Measurements\label{sec:zeno_prod}}

Now that we have discussed the already well-understood case of continuous-time quantum walk and adiabatic evolution, let us consider extending the previous work in \cite{Childs2002measurement}. As was discussed in that work, a projective measurement between the two lowest eigenstates of a search Hamiltonian can be performed in a time which is proportional to the inverse of the instantaneous gap between the ground and first-excited state of that Hamiltonian. This was argued by considering a model of measurement involving a continuous variable which, when compiled to a gate-model quantum computer, becomes the well-known phase estimation algorithm. We discuss the details of this model later in section \ref{sec:zeno_diss} where we use it to derive a physically realistic description of dissipation.

Since we are examining repeated projective measurements, it is easiest to consider this case using the density operator formalism. The action of measuring within a basis is to remove the off-diagonal elements of the density matrix within that basis, defined by $\ket{g}$ and $\ket{e}$. For a density operator $\rho$ the action of this quantum channel takes the form
\begin{align}
    \rho \rightarrow \Pi_{g}\rho \Pi_{g}+\Pi_{e}\rho\Pi_{e}= \nonumber\\ \ketbra{g}{g}\rho\ketbra{g}{g}+\ketbra{e}{e}\rho\ketbra{e}{e}=\ketbra{g}{g}\sandwich{g}{\rho}{g}+\ketbra{e}{e}\sandwich{e}{\rho}{e}.
    \label{eq:proj_measure}
\end{align}
As was observed in \cite{Childs2002measurement}, if we measure exactly at the avoided crossing where $\ket{g}=\frac{1}{\sqrt{2}}(\ket{\tilde{\omega}}+\ket{m})$ and $\ket{e}=\frac{1}{\sqrt{2}}(\ket{\tilde{\omega}}-\ket{m})$, the result of a single measurement is $\frac{1}{2}\left(\ketbra{g}{g}+\ketbra{e}{e}\right)=\frac{1}{2}\left(\ketbra{\tilde{\omega}}{\tilde{\omega}}+\ketbra{m}{m}\right)$. Therefore, a measurement in this basis will result in a $50$\% chance of finding the marked state. 

It follows from the structure of the eigenstates that measuring anywhere near the avoided crossing will result in a non-zero probability of $\ket{m}$. Moreover, performing a projective measurement can only decrease the purity of a density matrix. As argued in appendix \ref{appendix:alg_op} the probability of being found in the marked state can be lower bounded in terms of the purity of the density matrix,

\begin{equation}
\sandwich{m}{\rho}{m} \ge \frac{1}{2}\left(1-\sqrt{1-2\left(1-\mathrm{Tr}\left[\rho^2\right]\right)}\right). \label{eq:pure_marked}
\end{equation}

Since a projective measurement can only decrease $\mathrm{Tr}\left[\rho^2\right]$ it cannot decrease the lower bound of $\sandwich{m}{\rho}{m}$. Therefore, unlike in the case of dynamical evolution, there can be no exact cancellation from sequential measurements.

As long as at least one measurement is applied in the vicinity of the avoided crossing in a way which decreases the purity significantly, a sequence of measurements defines a valid algorithm  which gives a quantum speedup. For the purpose of this paper, the most interesting of these are measurements performed at sequentially decreasing values of $f(\tau)$ in a way which manifests as a (discrete) Zeno effect. Performing a large but not strictly infinite number of measurements yields a valid algorithm, however this will not have a well-defined continuum limit, since the total runtime will approach infinity. To have a well-defined continuum limit, we must define an operation which behaves like a projective measurement when performed for a time proportional to the gap, but is also meaningful for shorter times. We argue in the next section that decoherence in the diagonal basis of the Hamiltonian is one such operation. 

\begin{figure}
    \centering
    \includegraphics[width=10 cm]{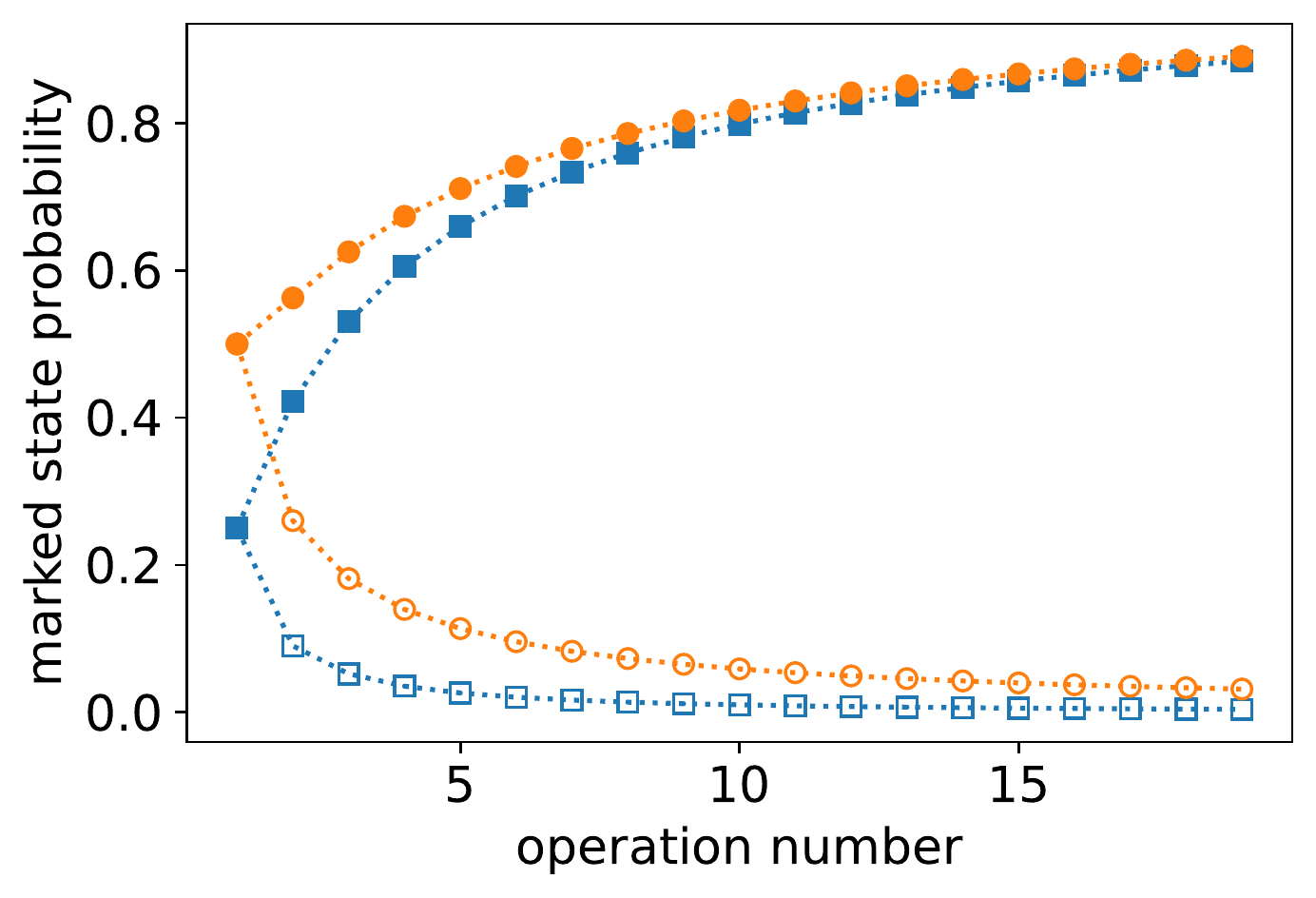}
    \caption{Marked state probability versus number of operations (either dephasing or destruction) performed (with placement determined by the optimal adiabatic schedule, as depicted in figure \ref{fig:sched_illustration}) for a variety of operations. Filled circles correspond to performing a number of projective measurements in the instantaneous energy eigenbasis of eq.~\ref{eq:H_ac} defined on the x-axis, while filled squares correspond to the same but partially destructive measurements. Unfilled symbols correspond to models (based on decoherence and dissipation, respectively) where the \textbf{total} rotation angles add to the fixed value for a single full operation. The filled symbols on the other hand correspond to versions where each operation corresponds to full dephasing (in the case of decoherence) or full removal of amplitude in the excited state (in the case of dissipation).}
    \label{fig:dephase_destroy_partial_full}
\end{figure}

An astute reader will notice that the arguments in the last paragraph do not strictly apply to a Zeno effect, since in this case the measurements do not significantly decrease the purity. We can however show numerically that a sequence of measurements can lead to a Zeno effect. Figure \ref{fig:dephase_destroy_partial_full} illustrates (as filled circles) that indeed if an increasing number of measurements are performed at points determined by the optimal adiabatic schedule, the marked state probability tends to unity. Moreover, we argue analytically that a Zeno effect of this form can effectively maintain the ground state in appendix \ref{appendix:alg_op}.

\subsection{Zeno Effect Induced by Decoherence \label{sec:zeno_dec}}

Decoherence can also lead to a Zeno effect, when subjected to pure decoherence (in the energy eigenbasis of a fixed Hamiltonian defined by ${\ket{g},\ket{e}}$) the density matrix will take the form
\begin{equation}
\rho=\rho_{gg}\ketbra{g}{g}+\rho_{ee}\ketbra{e}{e}+\rho_{eg}u(t_\mathrm{scale})\ketbra{g}{e}+\rho_{ge}u(t_\mathrm{scale})\ketbra{e}{g}, \label{eq:gen_decohere}
\end{equation}
where $u(\tau)\le 1$ and this is the term where the decoherence is manifested. Since decoherence decreases $\mathrm{Tr}\left[\rho^2\right]$, then by eq.~\ref{eq:pure_marked}, any protocol containing a single application of decoherence in the vicinity of the avoided crossing where $u(t_\mathrm{scale})$ becomes significantly less than one will lead to an optimal algorithm. 

\begin{figure}
    \centering
    \includegraphics[width=10 cm]{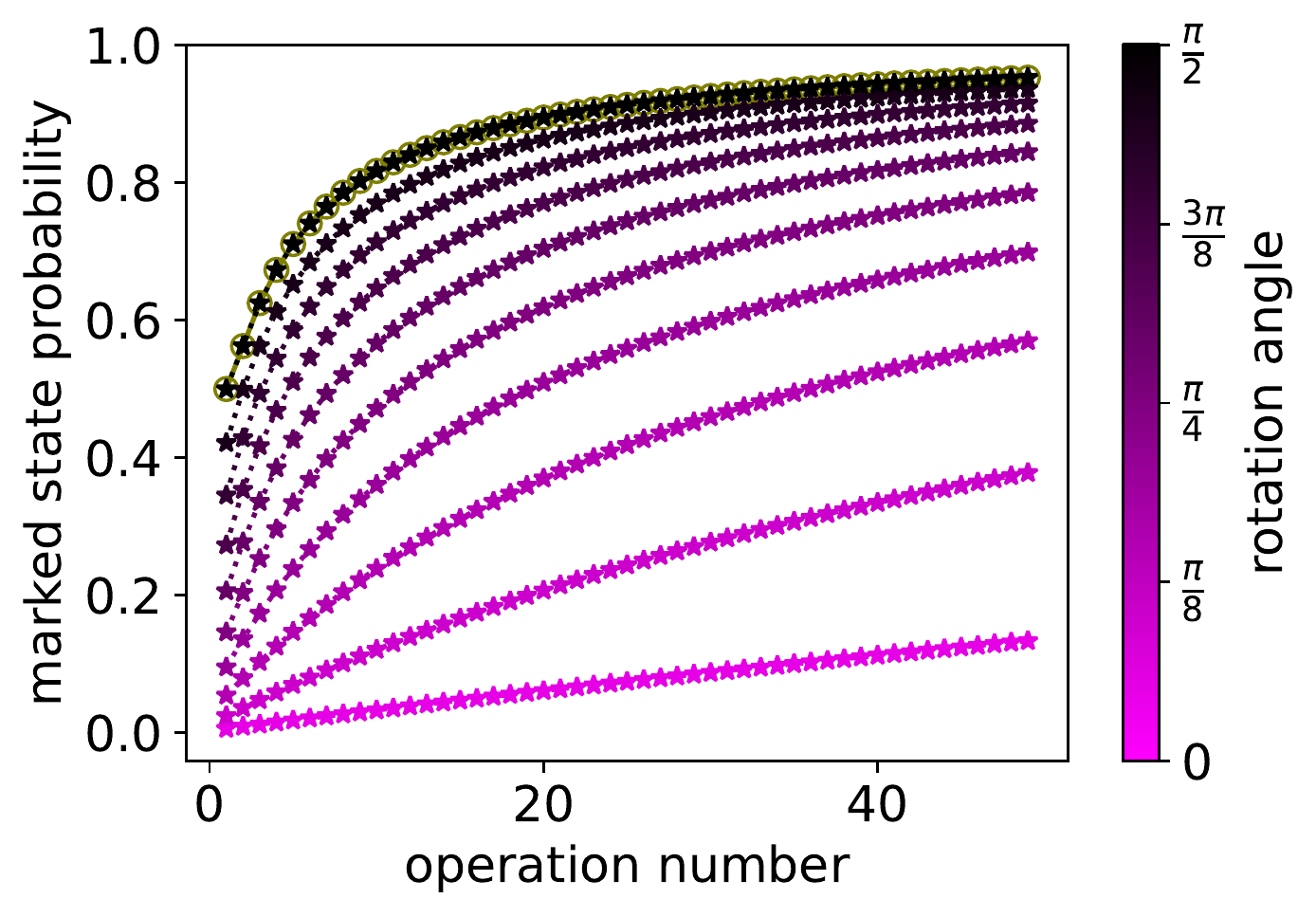}
    \caption{Marked state probability versus number of dephasing operations with different rotation angles per operation (encoded as colour/grayscale as depicted on the bar). Circles (gold) are used to indicate the projective measurement case. The values of $\tau$ for which the operations are performed are chosen to match the optimal adiabatic schedule.}
    \label{fig:dephase_meas_approach}
\end{figure}

This is getting ahead of ourselves a bit however, we first need to show that we can apply decoherence in a way which depends only on rescaled time, in other words with a timescale proportional to $g^{-1}_\mathrm{min}$. Starting with applications at fixed $f(\tau)=\gamma$, and then extending to multiple applications of decoherence. There will be multiple ways to show this, but we elect to build on results we already have for quantum walk. Consider a case where instead of just evolving with the Hamiltonian in equation \ref{eq:quantum_walk}, we instead flip a coin, if it is heads we apply 
\begin{equation}
 \exp\left(\frac{-i t_\mathrm{scale}}{2}\left[\gamma Z -X\right]\right),   
\end{equation}
but on tails we instead apply
\begin{equation}
 \exp\left(\frac{i t_\mathrm{scale}}{2}\left[\gamma Z -X\right]\right).  
\end{equation}
Summing the density matrices resulting from these two classical choices yields decoherence of the form \ref{eq:gen_decohere} with
\begin{equation}
u(t_\mathrm{scale},\gamma)=u(0)\cos\left(\sqrt{\gamma^2+1}t_\mathrm{scale} \right).
\end{equation}
Figure \ref{fig:dephase_meas_approach} shows the approach of the probabilistic quantum channel described here to the behaviour of a projective measurement. We define a $t_\mathrm{scale}$ which 
depends on $\tau$,
\begin{equation}
t_\mathrm{scale}=\frac{\phi}{\sqrt{\gamma^2+1}}, \label{eq:fixed_angle}
\end{equation}
this dependence means that for any value of $\tau$ there is a fixed rotation angle $\phi$, where $\sqrt{\gamma^2+1}$ is the rescaled energy gap between the ground and first excited state. In particular, total dephasing, with $\phi=\pi/2$ is identical to projective measurement. For simplicity, we initially focus on a constant $f(\tau)=\gamma$.

The simplest way (which as we show does not work) one could consider taking this model into a continuum limit is to consider applying decoherence of this form for an infinite number of vanishingly small periods of time. To do this we first expand the action of the decoherence over a short time period
\begin{equation}
u(\delta t)=u(0)-\frac{u(0)}{2}\left|\gamma^2+1\right|(\delta t)^2+O((\delta t)^4).
\end{equation}
Focusing for the moment on a fixed value of $f(\tau)=\gamma$, taking the limit in this way leads to
\begin{equation}
u_\mathrm{fin}=u(0)\lim_{m\rightarrow \infty} \left(1-\frac{\gamma^2+1}{m^2}\right)^m=\lim_{m\rightarrow \infty}u(0)\exp\left( \frac{\gamma^2+1}{m}\right)=u(0).
\end{equation}
In other words, in the continuum limit, the decoherence vanishes if defined in this way. The approach to this limit can be seen in figure \ref{fig:dephase_destroy_partial_full} (illustrated by the hollow circles). The reason for the somewhat counter-intuitive behaviour of this limit is actually an unintended Zeno effect, if we consider the effect of reversing the time evolution within the single avoided crossing model, this is equivalent to applying an operation to switch the role of the ground and first excited state, and then applying another one at the end. If we consider a specific set of choices to reverse or not reverse the time evolution at each time step $r_j\in \{0,1\}$, we have (building on eq.~\ref{eq:trotter_evolution})
\begin{align}
\ket{\psi_r(t)}&=\tilde{X}^{\sum_jr_j}\mathcal{T}\prod^{m}_{j=1}\exp\left(\frac{-i t_\mathrm{scale}}{2 m}\left[\gamma_j Z -X\right]\right)\tilde{X}^{r_j}\ket{\psi(t=0)} \nonumber\\&= 
\tilde{X}^{\sum_jr_j}\mathcal{T}\prod^{m}_{j=1}\exp\left(\frac{-i t_{\mathrm{scale}} \sqrt{\gamma^2_j+1}}{m}\tilde{Z}\right)\tilde{X}^{r_j}\ket{\psi(t=0)}, \label{eq:dephase_evolve} 
\end{align}
where $\tilde{X}=\ketbra{g}{e}+\ketbra{e}{g}$ if we take $\tilde{X}$ to an even power, including $0$ we obtain an identity matrix, terms within the product serve to reverse the time evolution, while the leading $\tilde{X}^{\sum_jr_j}$ term restores the correct ground and excited state in the case of an odd number of reversals. If we consider a uniform $50\%$ probability of a flip occurring at each time for the vast majority of choices of $r$, $\tilde{X}$ operations will occur frequently, with long ``runs'' of zeros in $r$ being exponentially rare. As $m\rightarrow \infty$, the dominant contribution will be cases where reversals are frequent. In this case a Zeno effect will occur, preventing relative phases from being acquired and therefore stopping all dynamics, leading to $\ket{\psi(t)}=\ket{\psi(0)}$, since the total density matrix can be defined as 
\begin{equation}
    \rho(t)=2^{-m}\sum_r\ketbra{\psi_r(t)}{\psi_r(t)}\rightarrow \ketbra{\psi(0)}{\psi(0)},
\end{equation}
this behaviour is illustrated numerically in figure \ref{fig:dephase_destroy_partial_full} by the unfilled circles.
A remedy to this issue is to consider a model where the $\tilde{X}$ operations become less probable within a given step proportional to $m^{-1}$, for example if we define using a source of uniform randomness $h_j\in \left[0,1\right]$
\begin{equation}
    r_j=\begin{cases} 1 & h_j < \frac{q}{m} \\ 0 &\mathrm{otherwise} \end{cases}.
\end{equation}
In this case, as long as $m>q$, then on average $q$ flips will occur, but this will not scale with $m$, so therefore we can set $q$ large enough to achieve a good approximation of continuous decoherence, but not strictly send it to infinity as we have to do to $m$ to achieve a continuum limit.

One way to define an effective coupling rate is to examine how much, on average, the state probability will transfer from the positive superposition state $(\ket{e}+\ket{g})/\sqrt{2}$ to the negative superposition state $(\ket{e}-\ket{g})/\sqrt{2}$ before the transfer direction is reversed. This expression divided by the time taken $\Delta p_\pm/\Delta t$ will give a quantity which can be roughly understood as ``effective'' coupling rate. This is coupling rate in the sense that the continuous process obtained by taking the limit $q\rightarrow \infty$ will effectively have a coupling rate equal to this quantity. Equation \ref{eq:dephase_evolve} implies that the effective coupling rate defined in this way (for a given $f(\tau)=\gamma$) is
\begin{align}
\kappa= \frac{\Delta p_\pm}{\Delta t}\approx \frac{q}{t_\mathrm{scale}}\left| \frac{\bra{e}-\bra{g}}{\sqrt{2}}\exp\left(\frac{-i t_{\mathrm{scale}} \sqrt{f^2(\tau)+1}}{q}\tilde{Z}\right)\frac{\ket{e}+\ket{g}}{\sqrt{2}} \right|^2 \nonumber \\
=\frac{q}{2t_{\mathrm{scale}}}(1-\cos\left(\sqrt{f^2(\tau)+1}\frac{2 t_\mathrm{scale}}{q} \right))\approx t_\mathrm{scale}\frac{f^2(\tau)+1}{2q}=\left(\frac{g(\tau)}{g_\mathrm{min}}\right)^2\kappa_0,
\end{align}
\begin{equation}
\kappa_0=\frac{t_z}{2}.
\end{equation}
Here we have effectively defined a Zeno time for our decoherence interaction \cite{Breuer2007open}, in our case the Zeno time for the decoherence inducing interaction is
\begin{equation}
t_z=\frac{t_{\mathrm{scale}}}{q} \label{eq:zeno_time}.
\end{equation}
The key aspect of this coupling for our purposes is that it is finite and independent of $g_\mathrm{min}$ and therefore the system size $N$. Continuous decoherence in a two state system can be described by a Lindblad equation with a Lindblad operator of $\tilde{Z}=\ketbra{g}{g}-\ketbra{e}{e}$, therefore this system will approximately obey the equation
\begin{equation}
    \frac{\partial}{\partial t}\rho=\kappa (\tilde{Z}\rho \tilde{Z}-\rho). \label{eq:dephase_lindblad}
\end{equation}
Where we note that the usual time evolution component of the Lindblad equation has vanished due to frequent reversal of the time evolution. This equation will become exact in the limit where $q \rightarrow \infty$, but is only nontrivial if $\kappa$ remains finite, which further implies that $t_\mathrm{scale} \propto q$ and therefore also $t_\mathrm{scale} \rightarrow \infty$. A more detailed derivation can be found in section \ref{appendix:eq:dephase_lindblad}. From here the arguments that this decoherence can lead to an optimal algorithm follow directly from the fact that the purity of the density matrix is reduced by the evolution, and therefore, from eq.~\ref{eq:pure_marked}, overlap with state $\ket{m}$ is unavoidable. 

\begin{figure}
    \centering
    \includegraphics[width=10 cm]{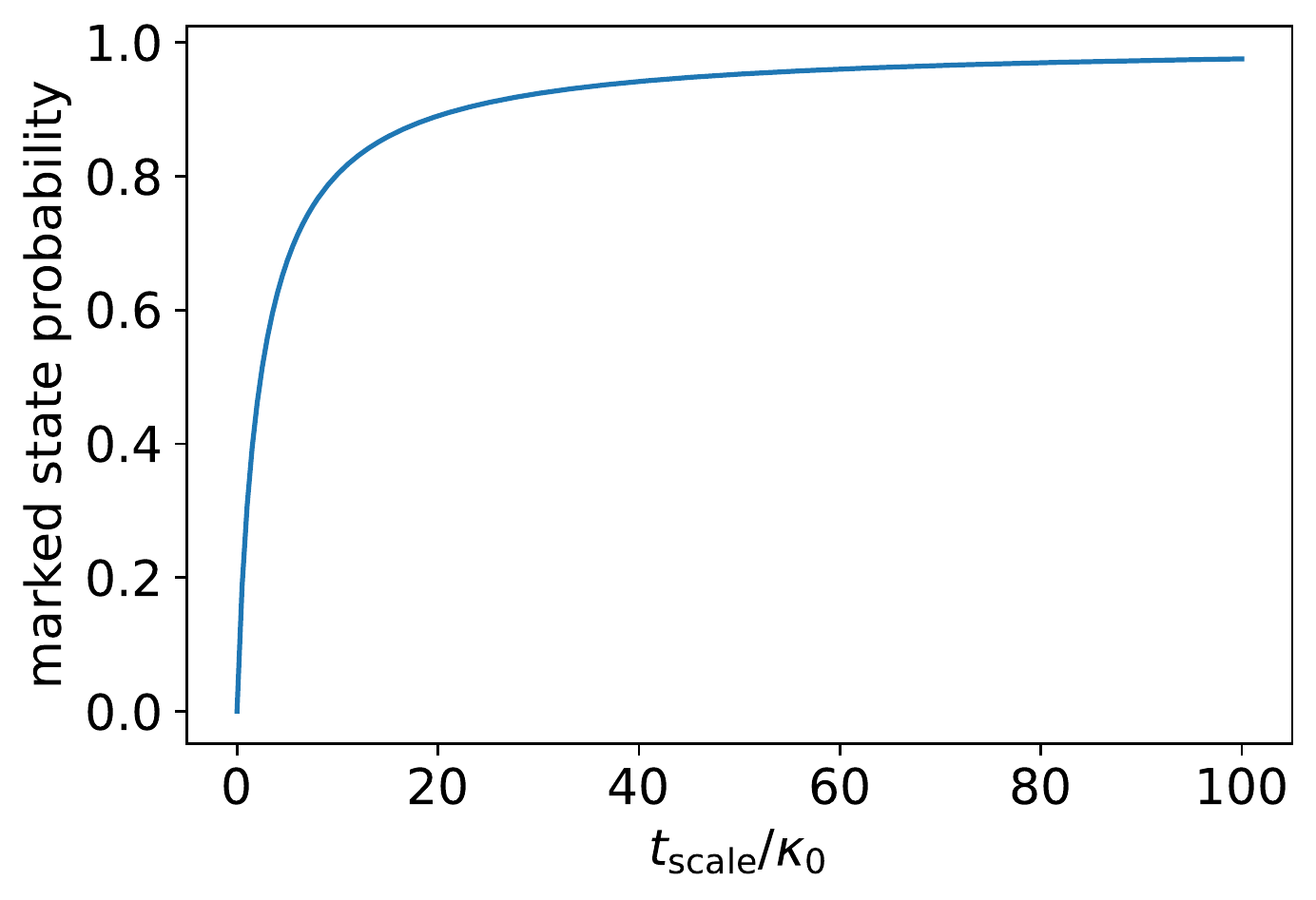}
    \caption{Probability of finding the marked state versus total scaled runtime for a Zeno effect due to continuous dissipation, modelled by equation \ref{eq:dephase_lindblad}.}
    \label{fig:cont_dephase_success_t}
\end{figure}

As before this argument does not strictly apply to cases where a Zeno effect is manifested, since these systems will remain in an approximately pure state throughout. We examine these cases numerically to show that indeed a Zeno effect can lead to finding the marked state with nearly $100\%$ probability.  Figure \ref{fig:cont_dephase_success_t} shows that a continuous Zeno effect does indeed lead to success probability which approaches $100\%$. We also show an alternative model for decoherence in appendix \ref{appendix:dec_model}.

\section{Destruction Family of Algorithms\label{sec:dest_family}}
\subsection{Zeno Effect from Partially Destructive Measurements}

Another kind of Zeno effect can be manifested due to dissipation, somewhat counter-intuitively a stronger dissipation can cause a Zeno effect which prevents the state which is coupled to dissipation from being occupied. As we did in the case of projective measurement and decoherence, we start with a model based on discrete measurements and then perform a continuum extension. To start with, we modify the model of measurement given in equation \ref{eq:proj_measure} to consider instead what we call a ``partially destructive'' measurement. When performing such a measurement, if the system is in $\ket{g}$ than it remains in $\ket{g}$, but if it is in state $\ket{e}$ it is sent into a ``destroyed'' state $\ket{d}$. The state $\ket{d}$ is considered computationally useless, if the system is in this state at the end of the computation, then the computation has failed and must be re-attempted. Perhaps unsurprisingly we will see that by taking advantage of the Zeno effect, protocols can be developed in which the probability of being found in $\ket{d}$ can be made arbitrarily small. In terms of density matrices, this operation takes the form
\begin{align}
    \rho \rightarrow \ketbra{g}{g}\rho\ketbra{g}{g}+\ketbra{d}{e}\rho\ketbra{e}{d}+\ketbra{d}{d}\rho\ketbra{d}{d}=\ketbra{g}{g}\sandwich{g}{\rho}{g}+\ketbra{d}{d}\left(\sandwich{e}{\rho}{e}+\sandwich{d}{\rho}{d}\right). \label{eq:dest_measure}
\end{align}
The fact that such an operation is possible in a time proportional to $\sqrt{N}$ follows directly from the arguments for projective measurements, since this just has an additional step of throwing away the state if a measurement takes a particular outcome.

The algorithm given in \cite{Childs2002measurement} will also work for this kind of measurement. A partially destructive measurement of this type right at the avoided crossing would result in the $\ket{d}$ state $50\%$ of the time, but would be found in the $\ket{g}$ state the other $50\%$, and of these cases a subsequent measurement would find $\ket{m}$ $50\%$ of the time, for an overall success probability of $25\%$ each time the algorithm is run. Since this is still $O(1)$ such an algorithm would still yield an optimal speedup. Similar arguments apply for measurements anywhere close to the avoided crossing. However, since amplitude in the excited state is ``lost'' this case does not have the elegant argument about purity which could be made in the projective case. In particular, a pair of sequential measurements where $|\gamma|$ takes the same large value, but has the opposite sign will result in the system being found completely in the $\ket{d}$ state since $\ket{e}$ with $\gamma$ would be $\ket{g}$ with $-\gamma$ and vice versa.

Zeno effects are however still possible, and indeed as the number of measurements are increased can approach $100\%$ success, as can be seen in figure \ref{fig:diss_meas_approach}. As with the case of projective measurement, there is no way to directly take a continuum limit of the partially destructive measurements. We instead need to take a continuous quantum channel which behaves like a partially destructive measurement, which as we discuss in the next section is dissipation out of the $\ket{e}$ state.

\subsection{Zeno Effect from Dissipation \label{sec:zeno_diss}}

As we did for decoherence, we must argue the existence of a physically realistic model which can dissipate some amplitude from $\ket{e}$ to $\ket{d}$ in a timeframe which scales as the amount of amplitude dissipated. To do this we build on the von Neumann model of measurement from \cite{Childs2002measurement}, but consider a different final treatment of the continuous variable, one which probabilistically causes dissipation if the system is in the $\ket{e}$ state, but does nothing if the system is in the $\ket{g}$ state. 

The model assumes coupling of the energy eigenbasis (defined by eigenvectors $\ket{E_a}$ and eigenvalues $E_a$) to a continuous variable through a $p$ operator, the general form is 
\begin{equation}
\sum_a\left[ \ketbra{E_a}{E_a}\exp\left(-it E_a p\right)\right] \label{eq:pCoup_general}
\end{equation}
The action of this Hamiltonian evolution is to send $\ket{E_a}\ket{x=0}$ to $\ket{E_a}\bigket{x=tE_a}$, wavepackets move distances apart corresponding to the energy differences multiplied by time. As was discussed in \cite{Childs2002measurement} this model can be compiled to a universal quantum computer, and in fact corresponds to the well known phase estimation algorithm in that case. In light of this fact, it is clear that this model can be considered physically realistic. A review of this model can be found in appendix \ref{appendix:comp_by_measure}.

Inserting our particular single avoided crossing model (and adding another qubit variable to the Hilbert space which will be used later in the algorithm) equation \ref{eq:pCoup_general} becomes 
\begin{align}
& U_\mathrm{meas}(t_\mathrm{scale},\tau) =  \nonumber \\
&\left[\ketbra{g}{g}\exp\left(it g_\mathrm{min}\sqrt{f^2(\tau)+1} p\right)+ \ketbra{e}{e}\exp\left(-it g_\mathrm{min}\sqrt{f^2(\tau)+1} p\right)\right]\otimes \mathbb{1}_2\nonumber \\ &=
\left[\ketbra{g}{g}\exp\left(it_\mathrm{scale}\sqrt{f^2(\tau)+1} p\right)+\ketbra{e}{e}\exp\left(-it_\mathrm{scale}\sqrt{f^2(\tau)+1} p\right)\right]\otimes \mathbb{1}_2,\label{eq:pCoup_ac}
\end{align}
where the purpose of the additional qubit subspace will become clear later.

As before, the scaling of the runtime as well as other quantities within the model removes all dependence on $g_\mathrm{min}$ and therefore $\sqrt{N}$, this indicates that as long as we can devise a scheme to perform an action which behaves like decay on the continuous variable, we can construct a model which is able to show an optimal speedup. Evolving from the state $(\psi_g\ket{g}+\psi_e\ket{e})\otimes\ket{x=0}$ with this Hamiltonian yields
\begin{equation}
    \ket{\psi_\mathrm{meas}}=\psi_g\ket{g}\otimes \bigket{x=t_\mathrm{scale}\sqrt{f^2(\tau)+1}}+\psi_e\ket{e}\otimes \bigket{x=-t_\mathrm{scale}\sqrt{f^2(\tau)+1}} \label{eq:pCoup_state}
\end{equation}

Let us now consider the additional single qubit which starts in state $\ket{0}$ can be evolved with the following Hamiltonian which couples it to the continuous variable 
\begin{equation}
    H_\mathrm{rot}=\mathbb{1}_2\otimes\left(\mathbb{1}\otimes X-x\otimes X\right)
\end{equation}
where $x$ is the position operator on the continuous variable and $X $ is the Pauli $X$ operator on the qubit. Applying evolution with this coupling to the state from equation \ref{eq:pCoup_state} gives
\begin{align}
U_\mathrm{rot}(t_\mathrm{rot})\ket{\psi_\mathrm{meas}}\otimes \ket{0}=\exp\left(-i t_\mathrm{rot} H_\mathrm{rot}\right)\ket{\psi_\mathrm{meas}}\otimes \ket{0}= \nonumber \\ \psi_g \ket{g}\bigket{x=x_\mathrm{mag}}\ket{0}+ \nonumber \\ \psi_e \ket{e}\bigket{x=-x_\mathrm{mag}}\left[\cos\left(2t_\mathrm{rot}x_\mathrm{mag}\right)\ket{0}-i\sin\left(2 t_\mathrm{rot}x_\mathrm{mag}\right)\ket{1}\right] \label{eq:rot_states}
\end{align}
where $x_\mathrm{mag}=t_\mathrm{scale}\sqrt{f^2(\tau)+1}$. To perform dissipation, we can then measure the qubit and abort the computation (equivalent to going to a useless state $\ket{d}$) if we measure in the state $\ket{1}$ (and continuing the computation if $\ket{0}$ is measured). As desired, this only happens if the system started in state $\ket{e}$, and then only with probability $\sin^2\left(2t_\mathrm{rot}x_\mathrm{mag}\right)$. The measurement operation is effectively a projection onto the zero state $\Pi_0=\mathbb{1}_2\otimes \mathbb{1} \otimes\ketbra{0}{0}$, this operation is non-unitary, but does allow us to continue to use an un-normalised version of the state vector representation, with the understanding that the ``missing'' amplitude represents the probability to be in state $\ket{d}$.

As a technical point, for this model to describe dissipation, it should not leave the system entangled with the continuous variable, so after measuring the qubit, we should apply the inverse of equation \ref{eq:pCoup_ac},
\begin{align}
    & U^\dagger_\mathrm{meas}(t_\mathrm{scale},\tau) =\nonumber \\
    &\ketbra{g}{g}\exp\left(-it_\mathrm{scale}\sqrt{f^2(\tau)+1} p\right)+\ketbra{e}{e}\exp\left(it_\mathrm{scale}\sqrt{f^2(\tau)+1} p\right),
\end{align}
Application of dissipation to $\ket{e}$ at a fixed value of $\tau$ therefore takes the form
\begin{align}
    \ket{\psi_\mathrm{dissap}}\otimes \ket{x=0} \otimes \ket{0}= \nonumber \\ D(\tau,t_\mathrm{scale},t_\mathrm{rot})\ket{\psi}\otimes \ket{x=0} \otimes \ket{0}= \nonumber \\
    U^\dagger_\mathrm{meas}(t_\mathrm{scale},\tau)\Pi_0 U_\mathrm{rot}(t_\mathrm{rot})U_\mathrm{meas}(t_\mathrm{scale},\tau)\ket{\psi}\otimes \ket{x=0} \otimes \ket{0}, \label{eq:disp_op}
\end{align}
where we note that all variables except for the original system variable are returned to their original state for the next application of dissipation. The model we have shown here can also be straightforwardly adapted to describe decoherence giving an alternative to the model in section \ref{sec:zeno_dec}, we show this construction explicitly in appendix \ref{appendix:dec_model}.

We have now shown that discrete steps which perform dissipation from the $\ket{e}$ to a useless state $\ket{d}$ are physically realistic. Numerical plots further show that these can give finite probability to be found in $\ket{m}$ and therefore yield an optimal speedup. 

\begin{figure}
    \centering
    \includegraphics[width=10 cm]{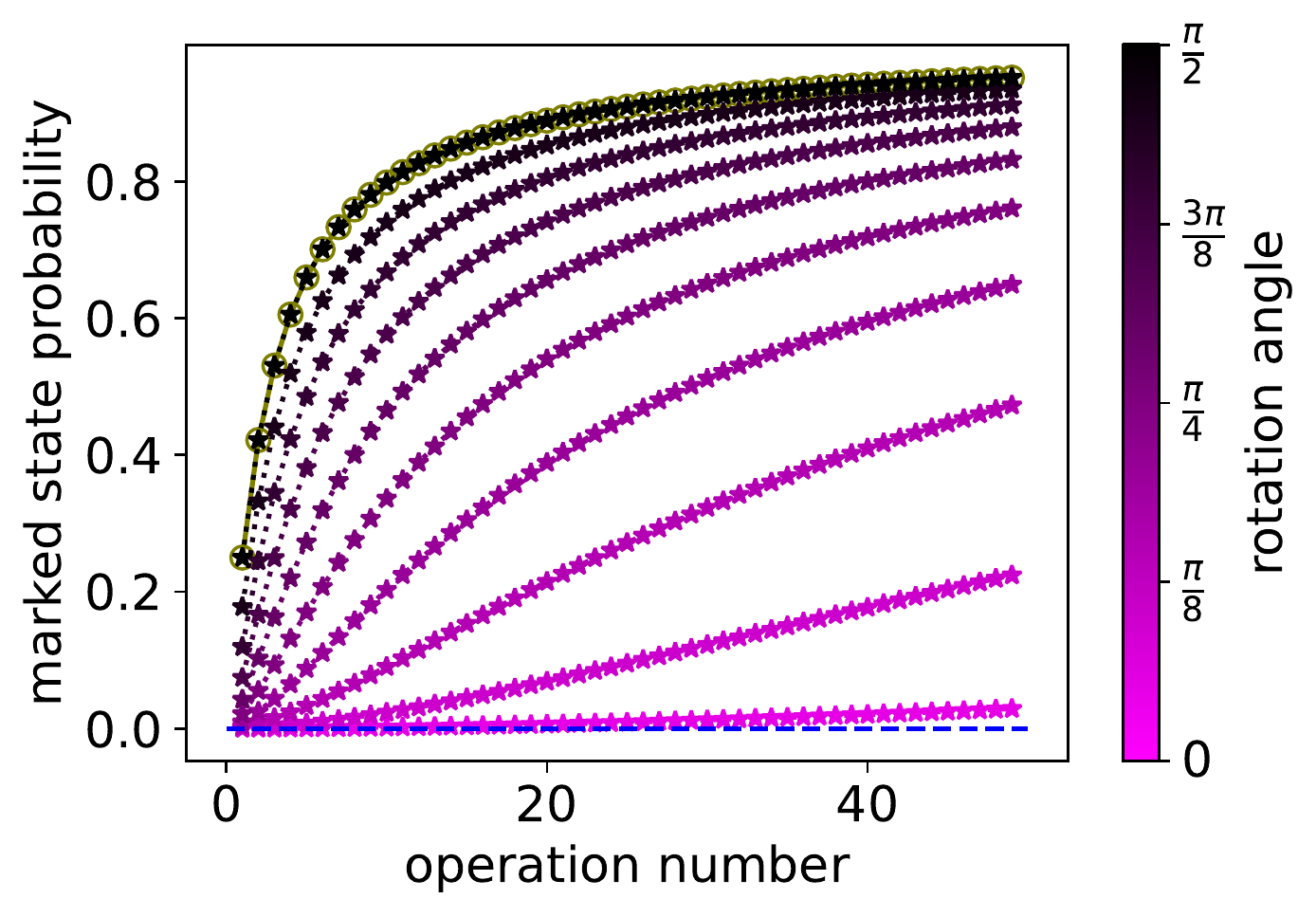}
    \caption{Marked state probability versus number of dissipation operations with different rotation angles per operation (encoded as colour/grayscale) as depicted on the bar. Circles (gold) are used to indicate the partially destructive measurement case. The values of $\tau$ for which the operations are performed are chosen to match the optimal adiabatic schedule. The dashed line is a guide to the eye at zero marked state probability to show that the lowermost curve is indeed (slowly) increasing.}
    \label{fig:diss_meas_approach}
\end{figure}

Equipped with this model, we can now perform multiple dissipative steps with a scaled rotation angle $\phi$, following equation \ref{eq:fixed_angle}, and setting $t_\mathrm{rot}=\pi/2$. The results can be found in figure \ref{fig:diss_meas_approach} and indeed show an approach to the behaviour of repeated partially destructive measurements. Before discussing the continuum limit, it is worth discussing a subtlety of this model.

Naively, from the equations we have written previously, it would appear that this dissipation model could yield a speedup which is even faster than the $\sqrt{N}$ Grover speedup. Namely, if we were able to take $t_\mathrm{rot}\propto N^\frac{1}{4}$ and $t_\mathrm{scale}\propto N^\frac{1}{4}$, we could obtain the same dynamics, but in a time which scales only as $N^\frac{1}{4}$. The reason that such an implementation is unphysical (it has to be, since it gives a scaling which is not possible in real quantum systems) is somewhat subtle. If we scaled in this way, $x_{mag}$ scales as $N^{-\frac{1}{4}}$. In such a scenario, the width of the initial state of the continuous variable which we have assumed are ``sufficiently narrow'' to be approximated as classical particles, would have to also scale down at least as $N^{-\frac{1}{4}}$. Such scaling is physically unrealistic, and without it, the final wavepackets would increasingly overlap as $N$ scales. For a large $N$ it would not be possible to reliably distinguish $\ket{e}$ from $\ket{g}$. 

In the actual model we have considered, the final difference in the continuum variable values does not scale with $N$, therefore the same ``sufficiently narrow'' width could work for all $N$ values. Similar arguments follow for all aspects of the interaction between the additional qubit and the variable, since all the scaling is confined between the interaction between the single avoided crossing system and the continuum variable.

We now consider what happens in the continuum limit. As with the decoherence case, we can see that as $t_\mathrm{scale}$ is decreased the probability of measuring in $\ket{d}$ after applying $D(\tau,t_\mathrm{scale}/q,t_\mathrm{rot})$ will scale as $1-\sin^2\left(2 t_\mathrm{rot}x_\mathrm{mag}\right)\propto t^2_\mathrm{scale}/q^2$. Therefore, if we divide the evolution into $q$ segments with $q\rightarrow \infty$, no dynamics will happen unless we also scale $t_\mathrm{scale}\propto q$. The reason for this is actually another Zeno effect. We are measuring the rotated qubit with increasing frequency, eventually stopping its dynamics. This behaviour is illustrated numerically in figure \ref{fig:dephase_destroy_partial_full} as hollow squares.

Fortunately, we can apply a similar technique as in the decoherence case to get a well-defined limit. Specifically, we note that we can perform the entire procedure in equation \ref{eq:disp_op}, but without the measurement $U^\dagger_\mathrm{meas}U_\mathrm{rot}U_\mathrm{meas}$, this operation will lead to qubit rotation, and leave it entangled with the avoided crossing system, but return the continuous variable to its original state. This strategy can allow amplitude to build up in the $\ket{1}$ state but avoids the unintended Zeno effect. Dissipation in the continuous time limit can be represented as
\begin{align}
&\ket{\psi(t)}  \otimes  \ket{x=0} \otimes \ket{0} = \nonumber \\
&\mathcal{T}\prod^q_{k=1}\Pi_0\mathcal{T}\prod_{j=1}^{m}U^\dagger_\mathrm{meas}(\frac{t_\mathrm{scale}}{mq},\tau_{k,j})U_\mathrm{rot}(t_\mathrm{rot})U_\mathrm{meas}(\frac{t_\mathrm{scale}}{mq},\tau_{k,j})\ket{\psi(0)}\otimes \ket{x=0} \otimes \ket{0},
\end{align}
where we can achieve continuum by taking\footnote{An astute reader may notice that in the previous formula $t_\mathrm{rot}$ is not scaled by $m$ and therefore strictly speaking the time grows with $m$, however we note that the time to complete $U_\mathrm{rot}$ is a factor of $\sqrt{N}$ smaller than $U_\mathrm{meas}$, so if the large $n$ where $N=d^n$ and large $m$ limits are taken simultaneously than this factor will vanish.} $m\rightarrow \infty$, but avoid Zeno effects by making $q$ large but not strictly infinite. Note that $\left[U^\dagger_\mathrm{meas},\Pi_0\right]=0$ due to the fact that they act on disjoint subspaces, this formula is equivalent to interspersing $D$ from equation \ref{eq:disp_op} into a sequence of operations without measurements.

\begin{figure}
    \centering
    \includegraphics[width=10 cm]{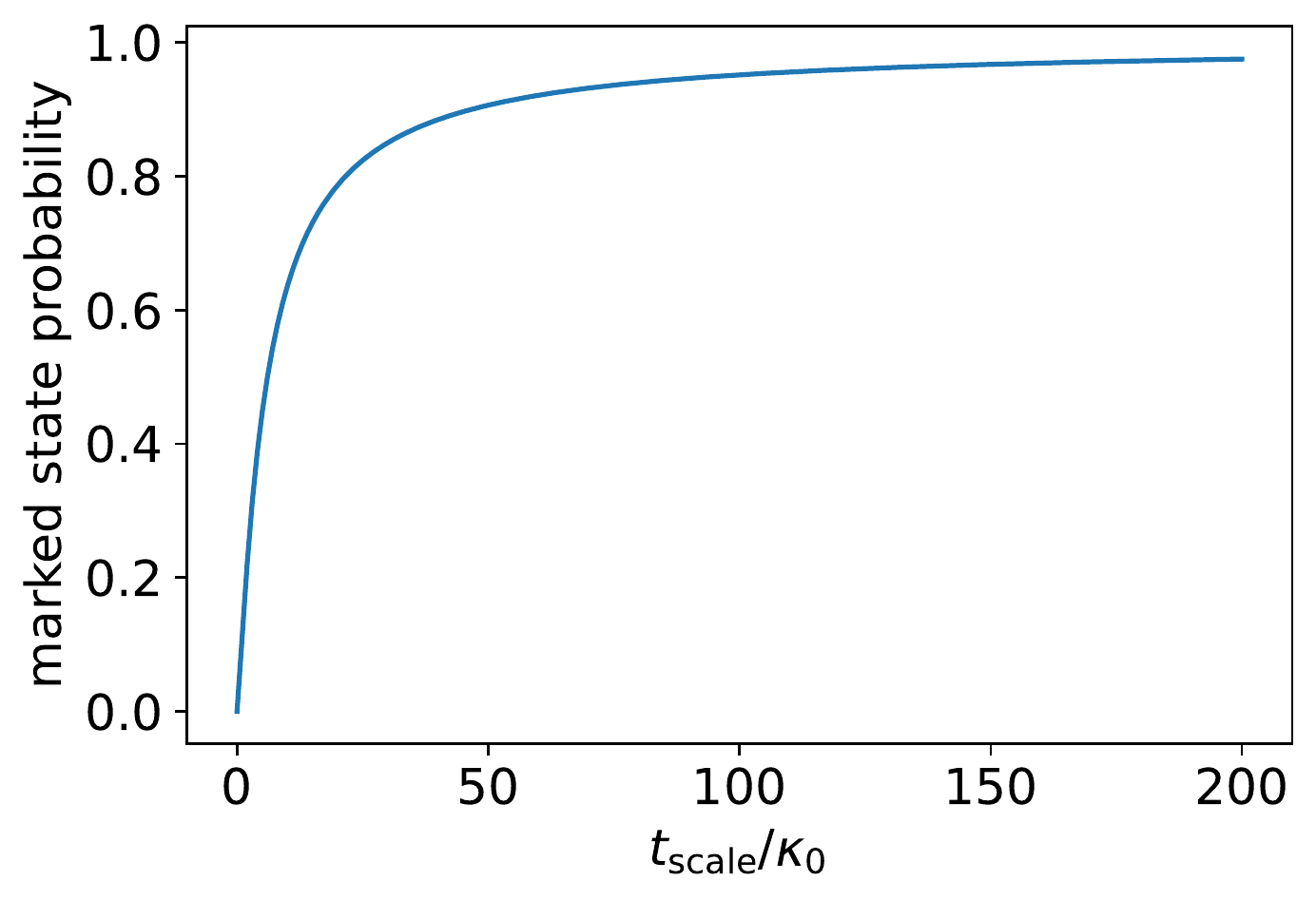}
    \caption{Probability of finding the marked state versus total scaled runtime for a Zeno effect due to continuous dissipation, modelled by equation \ref{eq:dissipate_lindblad}.}
    \label{fig:cont_dissipate_success_t}
\end{figure}

In the limit described above, infinite $m$ and large but finite $q$, each measurement effectively describes a fraction of the probability to be in state $\ket{e}$ being moved to $\ket{d}$ (represented by aborting the computation in the event of a $\ket{1}$ measurement). For a single measurement, the probability of this happening is $\sin^2\left(2 t_\mathrm{rot}x_\mathrm{mag}\right)$ according to equation \ref{eq:rot_states}. For large $q$ this can be converted into a decay rate by dividing by time
\begin{equation}
    \kappa\approx\frac{q}{t_\mathrm{scale}}\sin^2\left(2 t_\mathrm{rot}x_\mathrm{mag}\right)\approx 4 \frac{\left(f^2(\tau)+1\right)t^2_\mathrm{rot}t_\mathrm{scale}}{q}=\left(\frac{g(\tau)}{g_\mathrm{min}}\right)^2\kappa_0,
\end{equation}
as before, we can define $\kappa_0$ in terms of the Zeno time from equation \ref{eq:zeno_time},
\begin{equation}
\kappa_0=4 t^2_\mathrm{rot}t_z.
\end{equation}
Since the action of sending $\ket{e}\rightarrow \ket{d}$ can be represented as a Lindblad master equation with a Lindblad operator of $L=\ketbra{d}{e}$, the overall differential equation in the continuum limit becomes (see derivation in appendix \ref{appendix:eq:dissipate_lindblad}
)
\begin{equation}
    \frac{\partial}{\partial t}\rho=\kappa\left[\ketbra{d}{e}\rho\ketbra{e}{d}-\frac{1}{2}\left(\ketbra{e}{e}\rho+\rho \ketbra{e}{e} \right)\right]. \label{eq:dissipate_lindblad}
\end{equation}
As all dependence on $N$ has been removed from this scaled differential equation, any evolution resulting in a finite overlap with $\ket{m}$ corresponds to an algorithm which yields an optimal speedup (note that previous arguments imply that a finite overlap is all that is necessary). This is illustrated in figure \ref{fig:cont_dissipate_success_t}.

It is worth remarking that the same steps could be followed, but where the measured qubit is reset to $\ket{0}$ and the computation continued rather than aborting upon measuring $\ket{1}$ this would correspond to a Zeno effect through weak measurement which would have identical action to the decoherence examined previously, this is demonstrated explicitly in appendix \ref{appendix:dec_model}.

\section{Numerical Methods}

All numerical calculations were performed using Python, making use of numpy \cite{numpy} for simple matrix operations, and scipy\cite{2020SciPy-NMeth} (in particular scipy.linalg.eigh) for the numerical matrix diagonalisation required for figure \ref{fig:hyp20_ac_illustration} (a symmetric subspace was used, so the Hamiltonian matrix was of size $21$x$21$ and therefore did not require sparse diagonalisation methods). For all single avoided crossing calculations, the exact expressions for eigenvectors and eigenvalues were used, and therefore numerical diagonalisation was not required. Matplotlib\cite{hunter2007matplotlib} was used extensively for plotting.

\section{Discussion \label{sec:discussion}}

We now step back for a moment to summarise what we have shown. In each of the three families of Zeno effects we have studied, we have found three manifestations of the effect, which can all yield optimal speedup for unstructured search. These are summarised in table \ref{tab:families}, including the underlying equations describing the dynamics we examine in this work. By virtue of the nature of the scaling arguments, this also implies that interpolations of the form given in \cite{Morley19a} will be possible for the continuous versions of both the decoherence and destruction families of algorithms.

\begin{table}
\begin{centering}
\begin{tabular}{|c|c|c|c|}
\hline 
 & phase rotation & decoherence & destruction\tabularnewline
\hline 
\hline 
\multirow{2}{*}{fd} & phase flip & proj. meas. & part. dest. meas.\tabularnewline 
\cline{2-4} \cline{3-4} \cline{4-4} 
 & \bigstrut $\tilde{Z}\rho\tilde{Z}$ & $\Pi_{e}\rho\Pi_{e}+\Pi_{g}\rho\Pi_{g}$ & $\Pi_{g}\rho\Pi_{g}$\tabularnewline
\hline 
\multirow{3}{*}{pd} & ctqw & dephasing & dissipation\tabularnewline
\cline{2-4} \cline{3-4} \cline{4-4} 
 & \bigstrut $U(t)\rho U^{\dagger}(t)$ & $\frac{1}{2}U(t)\rho U^{\dagger}(t)+$ & $\cos^{2}(\frac{1}{2}g(\tau)t)\rho+$\tabularnewline
 & \bigstrut & $\frac{1}{2}U(-t)\rho U^{\dagger}(-t)$ & $\sin^{2}(\frac{1}{2}g(\tau)t)\Pi_{g}\rho\Pi_{g}$\tabularnewline
\hline 
\multirow{3}{*}{cont} & adiabatic & dephasing & dissipation\tabularnewline
\cline{2-4} \cline{3-4} \cline{4-4} 
 & \bigstrut $\frac{\partial}{\partial t}\rho=$ & $\frac{\partial}{\partial t}\rho=$ & $\frac{\partial}{\partial t}\rho=$\tabularnewline
 & \bigstrut $i\frac{g(\tau)}{2}\left[\rho\tilde{Z}-\tilde{Z}\rho\right]$ & $\kappa(\tau)\left[\tilde{Z}\rho\tilde{Z}-\rho\right]$ & $\kappa(\tau)\left[\Pi_{de}\rho\Pi_{ed}-\frac{1}{2}(\Pi_{e}\rho+\rho\Pi_{e})\right]$\tabularnewline
\hline 
\end{tabular}
\par
\end{centering}
\caption{Equations generating manifestations of different families of Zeno effects. In this work, we have shown that all manifestations given here are capable of yielding an optimal Grover speedup on unstructured search. To fit the table within the page width we have used some shorthand $U(t)=\exp(-ig(\tau)/2\tilde{Z})$, $\Pi_{de}=\ketbra{d}{e}$, $\Pi_{ed}=\Pi^\dagger_{de}=\ketbra{e}{d}$. In the leftmost column, fd is shorthand for ``full discrete'', pd for ``partial discrete'', and cont for ``continuous time''. We have also shortened ``continuous time quantum walk'' to ctqw. As a reminder of terms we have defined elsewhere in the paper $\tilde{Z}=\ketbra{g}{g}-\ketbra{e}{e}$,  $\Pi_{e}=\ketbra{e}{e}$, $\Pi_{g}=\ketbra{g}{g}$. Note that the first two entries for the destruction family do not conserve probability, it is implied that the ``missing'' probability is the probability to be in the computationally useless state $\ket{d}$.\label{tab:families}}

\end{table}

It is also worth remarking that, although not our focus, all nine combinations of families and manifestations will have algorithms with optimal scaling but which do not illustrate Zeno effects, examples which already exist in the literature are the search by measurement from \cite{Childs2002measurement} and a single instance of continuous-time quantum walk \cite{Childs2004spatial}. The interpolations from \cite{Morley19a} can behave highly non-adiabatically (and therefore not exhibit a Zeno effect), but are still based on continuous time evolution like adiabatic quantum computing and therefore belong to the phase rotation family of algorithms. While we will not examine it further, by drawing parallels between the models, it is clear that similar interpolations will be possible for continuous decoherence and dissipation.

Given the breadth of algorithms we have chosen to study in this work, and in the interest of length, we have not examined the effect of finite size on our algorithm as was done in \cite{Morley19a}. We expect these would have similarly rich structure to what was found in that work, and this is likely a fruitful area of future research.

Comparing figures \ref{fig:msqw_mult_angle}, \ref{fig:dephase_meas_approach}, and \ref{fig:diss_meas_approach} in the discrete case and \ref{fig:adiabat_success_t}, \ref{fig:cont_dephase_success_t}, and \ref{fig:cont_dissipate_success_t} in the continuous case, we observe a pattern in the constant factors on the scaling for our very simplified model. We find that the phase rotation family has the most favourable constant factors, followed by decoherence, followed by destruction. However, two remarks should be made here, firstly it is not clear whether or not these factors would translate to structured (optimisation) problems. Secondly, even if they did, if it were easier to implement decoherence or destruction manifestations of a Zeno effect (see for example \cite{nguyen2024entropycomputing}), then it could still be preferable to apply those algorithms in a real setting.

We have further not studied how the mechanisms interact with each other. While significant work has been put into understanding the effect of open-system effects in quantum annealing \cite{Thurnstrom2005open,Albash2012open,Sarandy2005open,Venuti16open}, we are not aware of work which focuses on strong interactions which cause a Zeno effect. A fuller understanding of how the mechanism we study here interact are likely to be an interesting area for future research.

While the results we have shown are interesting from a theoretical perspective, a question we have not fully addressed thus far is what practical relevance the models we have discussed here have. In particular, while theoretically useful, all the Hamiltonians we have discussed here would be more complicated to implement than directly implementing an unstructured search Hamiltonian (for example using the methods of \cite{Dodds2019permutation}). A natural question is whether there is actually a physical situation where it may be easier to implement either a dephasing or destruction model. We discuss at least one technology which answers this question in the affirmative for the destruction model. Note that we will not discuss how to actually encode the problem to this paradigm, as that is beyond the scope of this current work.

In section \ref{sec:intro} we previewed the potential relevance of this work to the entropy computing paradigm through quantum Zeno blockade. The realisation of a Zeno blockade involves an optical cavity (for example a micro-ring resonator) which contains or is made of a non-linear optical material which supports sum-frequency generation \cite{Sun2013Blockade}. There also has to be a freely propagating mode which starts outside the cavity (for example, an optical fibre coupled to the cavity). The Hamiltonian of the simplest model of the Zeno blockade (ignoring the photon propagation in the fibre as a simplification for illustrative purposes) is therefore\footnote{Here we use standard quantum optics notation, where all operators are given hats, including Hamiltonians}:

\begin{equation}
    \hat{H}_\mathrm{ZB}=c\left(\hat{a}_{f}\hat{a}^\dagger_\mathrm{cav}+\hat{a}^\dagger_{f}\hat{a}_\mathrm{cav}\right) +G \left(\hat{a}_{cav}\hat{a}_\mathrm{cont}\hat{a}^\dagger_\mathrm{loss} + \hat{a}_\mathrm{cav}^\dagger\hat{a}^\dagger_{cont}\hat{a}_\mathrm{loss}\right)
\end{equation}

From an engineering perspective, it is difficult to make a cavity which is highly resonant with three modes at once \cite{Chen2017Blockade,Ma2020Ultrabright}. As a result, it is natural to also consider loss from one of the modes (the mode corresponding to the operator $\hat{a}_\mathrm{loss}$). This results in a total master equation of the form

\begin{equation}
\frac{\partial \hat{\rho}}{\partial t}=-i[\hat{H}_\mathrm{ZB},\hat{\rho}]+\gamma \left[\hat{a}_\mathrm{loss}\hat{\rho} \hat{a}^\dagger_\mathrm{loss}-\frac{1}{2}(\hat{a}^\dagger_\mathrm{loss}\hat{a}_\mathrm{loss}\hat{\rho}+\hat{\rho}\hat{a}^\dagger_\mathrm{loss}\hat{a}_\mathrm{loss})\right]. \label{eq:ZB_master}
\end{equation}

In this setting, the Zeno effect is modulated by the off-diagonal density matrix element which captures the coherence between a freely propagating mode (corresponding to $\hat{a}_f$) and the captured cavity mode (corresponding to $\hat{a}_\mathrm{cav}$). In fact, only the imaginary component of this term contributes to transfer of a freely propagating photon into the cavity, this leads to (assuming $m$ photons are present in the control mode and $n$ in the freely propagating mode),
\begin{equation}
\frac{\partial}{\partial t}\mathrm{Im}\left[\rho_{cf}\right]=-\sqrt{n}c \left( \rho_{ff}-\rho_{cc}\right)-\sqrt{m\,n}\,G\,\mathrm{Re}\left[\rho_{lf}\right].\label{eq:Im_rho_cf_diff}
\end{equation}
For the density matrix element $\mathrm{Re}\left[\rho_{lf}\right]$, we have the formula
\begin{equation}
\frac{\partial}{\partial t}\mathrm{Re}\left[\rho_{lf}\right]=\sqrt{m\,n}\,G\,\mathrm{Im}\left[\rho_{cf}\right]+\sqrt{n}\,c\,\mathrm{Im}\left[\rho_{lc}\right]-\frac{\gamma}{2}\mathrm{Re}\left[\rho_{lf}\right]. \label{eq:Re_rho_lf_diff}
\end{equation}
If we assume that an effective Zeno blockade is implemented, then terms proportional to $c$ can be neglected, leading to an overall combined differential equation of
\begin{equation}
\frac{\partial^2}{\partial t^2}\mathrm{Im}\left[\rho_{cf}\right]= -m\,n\,G^2\mathrm{Im}\left[\rho_{cf}\right]-\frac{\gamma}{2}\frac{\partial}{\partial t}\mathrm{Im}\left[\rho_{cf}\right], \label{eq:off_diag_diff}
\end{equation}
which is the same underlying differential equation as a damped harmonic oscillator. The term proportional to the loss rate $\gamma$ acts as an effective damping term, When $4\,\sqrt{n\,m}\, G>\gamma$, this equation is effectively under damped and $\mathrm{Im}\left[\rho_{cf}\right]$ will oscillate between positive and negative values, as it decays, characteristic in a phase rotation manifestation of the Zeno effect. On the other hand if $4 \,\sqrt{n\,m}\,G\le \gamma$ the system is over (or critically) damped then $\mathrm{Im}\left[\rho_{cf}\right]$ will decay but not change sign, as would occur in a destruction based manifestation of the Zeno effect. These regimes roughly map to the coherent (CQZ) and incoherent (IQZ) quantum Zeno blockade regimes as discussed in previous literature \cite{huang2012antibunched,Sun2013Blockade,Huang2010Switch} This regime is also the one which is more likely to be realised with small $m$ and $n$, in particular in the $m=n=1$ case of a single-photon Zeno blockade. This few or single photon regime is the one which is most relevant to entropy computing \cite{nguyen2024entropycomputing}.

\section{Conclusion}

First and foremost we have demonstrated that within a physically realistic setting, quantum Zeno effects due to decoherence (or equivalently measurement) and decay into a computationally useless state, are able to support an optimal quantum speedup for unstructured search. This is an important step because it demonstrates that analog optimisers, in the spirit of quantum annealers, but using these effects to achieve a Zeno effect are well motivated. This is not a given, since these quantum channels are often considered undesirable and to detract from the quantum nature of a device. Discussing how such a device could actually be achieved is beyond the scope of this work, as our current focus is on what is theoretically achievable. 

It is important to note that this extends beyond previous work which has shown that for example universal gate-model quantum computation can be based on Zeno-effect gates \cite{Huang2008zenouniversal}. In these settings, the effects would typically be limited to one or two qubit subspaces. What we have shown is that analog quantum computation where a many-body manifestations of the effects are present can also support a quantum advantage. Such specialized devices are likely to be easier to build in the near term, in analogy to the setting of quantum annealing.

We have furthermore catalogued a wide variety of potential algorithms which can realise an optimal speedup, some of which represent manifestations of Zeno effects and some of which do not. We have organized these in terms of category and mode of operation, whether they operate discretely or continuously, and if discrete whether they are based on full operations which maximally disturb the system (such as projective measurements) or not. This organised structure allows a high-level understanding of the relationship between these algorithms, in the same spirit of the interpolation between adiabatic computing and continuous time quantum walks in \cite{Morley19a}. While we have not shown it explicitly in this work, it is worth noting that similar interpolations will be possible for the continuous manifestations of the other two families of algorithms as well.

The intention of this work is to build a foundation upon which new directions in analog quantum computing devices can be explored, equipped with the guarantee that a broad class of methods can directly support a speedup.

\section{Acknowledgements}

All authors were fully supported by Quantum Computing Inc.~in completing this work. The authors thank Milan Begliarbekov
and Yong Meng Sua for useful discussions and references, and Uchenna Chukwu for spotting some small errors in the text. 

\section*{Appendices}
\appendix

\section[Review of model]{Review of the model from \cite{Childs2002measurement} \label{appendix:comp_by_measure}}

Since the work in our present manuscript builds on a model originally presented in \cite{Childs2002measurement} it is worth reviewing the key aspects of the model of measurement from that paper, to make our paper more self-contained. It is worth noting that \cite{Childs2002measurement} briefly mentioned Zeno based computation by measurement early in the paper, but does not analyse the potential for a quantum speedup. In fact, the protocol they discuss is one based on linear interpolation, which would not yield a quantum speedup for search \cite{Roland2002}.

The model we are interested in was rather used to motivate the amount of time over which a single measurement could be performed. This time is shown to scale as $\sqrt{N}$. The measurement time combined with the fact that after this measurement there is an $O(1)$ success probability independent of system size, implies that an algorithm based on single measurements is theoretically able to achieve the same scaling as Grover's famous gate model algorithm \cite{Grover1996search,Grover1997Search}. A key fact to note is that the model of measurement from \cite{Childs2002measurement} is not necessarily meant to replicate a real physical device, so much as to show that it is physically realistic to be able to measure on a timescale proportional to $\sqrt{N}$. Our argument follow a similar line of reasoning, to explore what is possible, rather than propose a concrete design. It is worth noting that the Hamiltonian used in this model can be viewed as a continuous-time variant of gate model phase estimation \cite{kitaev1995quantummeasurement,cleve1998phaseEstimate}. This connection was made explicit in section 4 of \cite{Zalka1998simulate}.

The key idea behind this model is that an arbitrary von Neumann measurement can be represented as time evolution under a Hamiltonian which couples as system to a continuous variable, which is then measured to obtain the value. Let us assume the goal is to measure the expectation value of a Hermitian operator $\hat{A}$. We will follow the notation of \cite{Zalka1998simulate}\footnote{While we do not use hats for operators elsewhere, it is useful to use them here for clarity and consistency with other literature, we do however retain the dimensionless units of time and energy, which effectively remove factors of $\hbar$.} and consider measuring an eigenstate of $\hat{A}$, $\ket{\Psi_a}$ with an eigenvalue of $a$. This leads to the property that 
\begin{equation}
\hat{A}\ket{\Psi_a}=a\ket{\Psi_a}.
\end{equation}
We further define a continuous system initialised at a single position, $\ket{x}$. As usual this system will have position and momentum operators $\hat{X}$ and $\hat{P}$ respectively, and these will obey the commutation relationship $\left[\hat{X},\hat{P}\right]=i$. A measurement can then be conceived of as effectively evolving with a Hamiltonian of the form 
\begin{equation}
\hat{H}=-k\hat{A}\hat{P}.
\end{equation}
Applying time evolution with respect to this Hamiltonian yields 
\begin{equation}
\hat{U}(t)\ket{\Psi_a}\ket{x}=\ket{\Psi_a}\ket{x+k\, a\, t}.
\end{equation}
By the linearity of quantum mechanics, applying this operator to a superposition of such eigenstates will lead to a superposition of position values. 

However, we must recall a key property of measurement of continuous systems is that the position cannot be measured with infinite precision (both because $x$ eigenstates cannot be prepared perfectly, and because of limitations of the apparatus, we consider the former in the main text for convenience). This precision is clearly independent of any properties of $\hat{A}$, or the runtime of the experiment. If we assume a fixed precision at which $x$ measurements can be reliably performed $\Delta x$, and our aim is to differentiate states with eigenvalues $a_0<a_1$, then the required time for the measurement will be 
\begin{equation}
    t=\frac{\Delta x}{k (a_1-a_0)},
\end{equation}
in other words, it will depend inversely on the difference in expectation values of $\hat{A}$, which is the energy gap in the case where $\hat{A}$ is a Hamiltonian\footnote{Note that as a Hamiltonian $\hat{A}$ would have a natural energy scale so a trivial rescaling of $\hat{A}\rightarrow c\hat{A}$ would not be allowed.}. 

It is worth briefly noting that the evolution described here can be simulated in a gate model setting. In this setting, a convenient way to implement the time evolution operator $\hat{U}(t)$ is in the momentum basis rather than the real space basis \cite{kitaev1995quantummeasurement,cleve1998phaseEstimate}. In this implementation, a register is initialised in $\ket{x=0}$ which in the momentum basis is an equal positive superposition of all bitstrings. Each bit is then used as a control on a small angle unitary of with rotation angles proportional to different binary numbers. This imprints phases on the binary-encoded momentum state. An inverse quantum Fourier transform then converts these into a binary representation of $x$ which can be measured. This implementation is the famed phase estimation algorithm, it is reviewed for example in section 5 of \cite{cleve1998phaseEstimate}, but we will not discuss the details here since this work focuses on the continuous-time setting.

\section{Unstructured Search Hamiltonian\label{appendix:unstruct_search}}

The original unstructured search algorithm by Grover was based on a series of sequential rotations, typically compiled to gate-model quantum computer \cite{Grover1996search,Grover1997Search}. In this setting, an ``oracle'' operation applies a phase to an unknown (to the person implementing the algorithm) marked state. The scaling is then measured as the number of times the oracle is applied, because that is the only step which requires awareness of which state is marked.

An alternative paradigm is to define an oracle Hamiltonian, \cite{Farhi1998QuantumWalk} which continuously applies a phase rotation to every state except for the marked state $\ket{m}$, this Hamiltonian takes the form:
\begin{equation}
H_{\mathrm{mark}}=\mathbb{1}-\ketbra{m}{m} = \sum_{j\neq m} \ketbra{j}{j},\label{eq:H_mark}
\end{equation}
it is worth briefly noting that since exponentiating $\mathbb{1}$ applies an unmeasureable global phase, the action of this Hamiltonian is equivalent to evolution with $H_{\mathrm{mark}}=-\ketbra{m}{m}$. The factor of $\mathbb{1}$ is included as a matter of convention. 

To apply computation in this setting, an additional ``driver'' Hamiltonian that doesn't commute with $H_{\mathrm{mark}}$ must be added to induce non-trivial dynamics, 
\begin{equation}
H(s)=(1-s) H_{\mathrm{driver}}+s H_{\mathrm{mark}}.
\end{equation}
Two drivers are commonly used here, the most mathematically simple version is a fully connected graph driver  
\begin{equation}
H_{\mathrm{driver}}=H_{\mathrm{fc}}=\mathbb{1}-\ketbra{\omega}{\omega},
\end{equation}
where 
\begin{equation}
    \ket{\omega}=\frac{1}{\sqrt{N}}\sum_i\ket{i},
\end{equation}
an equal positive sum over all computational basis states.
Such a driver allows direct transitions from any computational basis state to any other computational basis state. The advantage of this driver is that the Hamiltonian
\begin{equation}
H_\mathrm{full}(s)=\mathbb{1}-(1-s)\ketbra{\omega}{\omega}-s\ketbra{m}{m} \label{eq:h_with_full}
\end{equation}
can be diagonalised by hand within a two-dimensional symmetric subspace spanned by $\{\ket{\omega},\ket{m}\}$ 
and therefore facilitates calculations. Since $|\braket{\omega}{m}|>0$ in all finite cases, we work in a space defined by the orthogonal vectors spanned by $\{\ket{\tilde{\omega}},\ket{m}\}$  by construction $\braket{\tilde{\omega}}{m}=0$. We also assume a starting state of $\ket{\tilde{\omega}}$, for most subsequent calculations. What this does is to effectively ignore the order $N^{-1}$ effect of ``randomly guessing'' the marked state. These terms become increasingly irrelevant as $N$ grows, and neglecting them makes the calculations significantly simpler.

A disadvantage of the fully connected driver Hamiltonian in equation \ref{eq:h_with_full} is that the driver describes a process which is extremely difficult to implement physically. Every computational basis state couples equally to every other computational basis state. Such a Hamiltonian is highly non-local in the underlying qubits (or qudits) used to implement an analog algorithm. 

A driver which is often considered due to the relative ease of engineering it physically is the transverse field driver.  
This driver can be described entirely using local Pauli $X$ operations $X=\left(\begin{array}{cc}0 & 1\\ 1 & 0 \end{array}\right)$,
\begin{equation}
H_{\mathrm{driver}}=H_{\mathrm{tf}}=-\sum_{i=1}^nX_i.\label{eq:H_trans}
\end{equation}
The minus sign is included to guarantee that the ground state of the Hamiltonian is $\ket{\omega}$. Since this driver induces single bit flips, the graph which this Hamiltonian forms within the solution space is a hypercube of dimension $n$, with the computational basis states as the corners of the hypercube and the edges of the hypercube representing the single bit flip operations. For this reason, especially within the quantum walk community where the underlying graph plays a central role, this driver is sometimes referred to as a hypercube hopping term. It is worth remarking that unstructured search using this driver can be engineered at second order in perturbation theory using only two-body Ising interactions \cite{Dodds2019permutation}.

A further extension is to qudits, where each variable has dimension $d$. Assume a driver which has equal hopping between adjacent states $\ket{i}\leftrightarrow \ket{i+1}$. The graph formed by this driver within the solution space is again a hypercube, but includes interior points, not just corners. In the context of quantum walks, it has been shown that a Grover speedup is possible for hypercubes of dimension four and greater ($n\ge 4$ in our notation), with an optimal speedup for $n>4$ \cite{Childs2004spatial} and $d\ge 2$. This structure for quantum walk implies a description with a single avoided crossing model which will be discussed in the next section, so our result extends to qudits as well, except for the case of three or fewer qudits. Computing with a fixed number of qudits does not represent a valid way to perform scalable quantum computing \cite{Blume-Kohout02a}, so these cases are not interesting for understanding computational advantage. 

\section{Single Avoided Crossing Model \label{appendix:avoided_cross}}

Although it is mathematically harder to show \cite{Childs2004spatial,Morley19a}, the relevant search dynamics when using a transverse field driver as given in Eq.~\ref{eq:H_trans}, become increasingly well described as a two level system, with an avoided crossing having a gap proportional to $N^{-1/2}=2^{-\frac{n}{2}}$, where the two-level system implies that $N=2^n$.  As discussed in the last section, this also extends to qudits for $n>4$, 
with a gap proportional to ${N}^{-1/2}=d^{-\frac{n}{2}}$, where $N=d^n$ in this case. For this reason we are able to consider a model which captures multiple cases, the single avoided crossing model from \cite{Morley19a}, which was shown in that paper to be valid for a transverse field driver as the size becomes asymptotically large, and by virtue of the two-level nature of the fully connected graph driver Hamiltonian, is also a valid description of that Hamiltonian in the large size limit. A similar approach, but without an explicit model, was also used in \cite{Chakraborty2016randGraphs}. Additionally, this structure is present in more general hypercubes which describe the solution space of qudit systems \cite{Childs2004spatial}. In the limit of large systems, this model can be expressed in the terms of the previously defined Pauli $X$ operators and the Pauli $Z$, $Z=\left( \begin{array}{cc}1 & 0 \\ 0 & -1 \end{array} \right) $. The approach to this limiting behaviour has been extensively studied in \cite{Morley19a} The generally time dependent Hamiltonian in the single avoided crossing approximation takes the form 
\begin{equation}
    H_{\mathrm{ac}}=\frac{g_{\mathrm{min}}}{2}\left[f(\frac{t}{T})Z-X\right],
\end{equation}
where $T$ is the total runtime and $g_{\mathrm{min}}$ is the minimum energy gap between the ground and first excited state. Since the schedule is defined in terms of the runtime as a fraction of the total rather than the absolute runtime, we define a rescaling $t/T\rightarrow \tau$ and use $\tau$ as the relevant time parameter.

\begin{figure}
    \centering
    \includegraphics[width=9 cm]{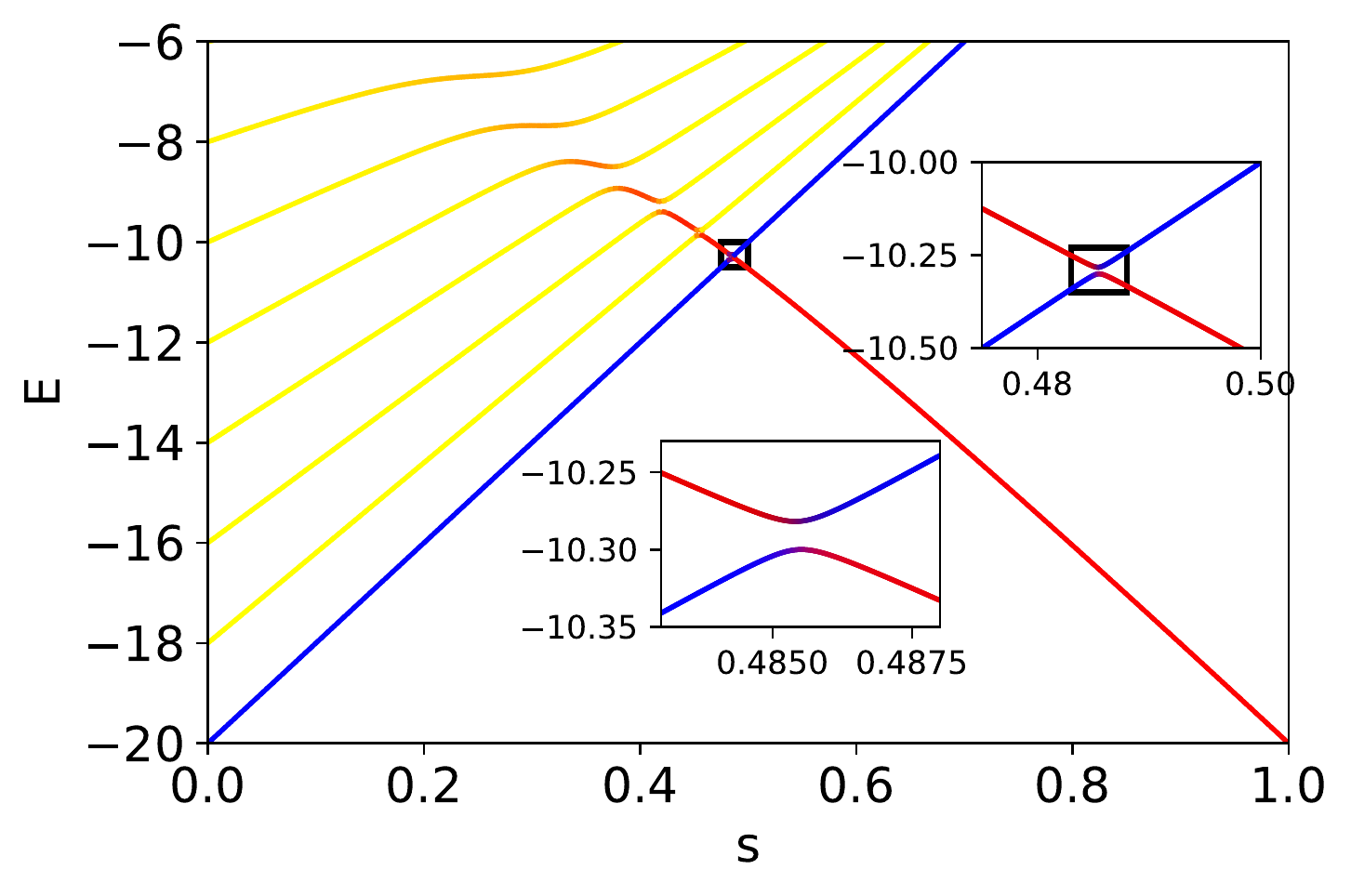}
    \caption{Approach to single avoided crossing behaviour for a hypercube search with $n=20$. The y-axis depicts the energy of the eigenstates, while the x-axis depicts the parameter $s$ for A Hamiltonian of the form $H=(1-s)H_\mathrm{tf}+n s \ketbra{m}{m}$. Here $H_\mathrm{tf}$ comes from equation \ref{eq:H_trans}. Insets are zooms of the same plot, with the right box showing a zoom within the box depicted on the main figure and the lower inset showing a zoom on the right plot. Color indicates eigenstate probability to be measured in relevant states, with red indicating the marked state and blue indicating $\ket{\tilde{\omega}}$; yellow indicates any other state.}
    \label{fig:hyp20_ac_illustration}
\end{figure}

The physical intuition behind this Hamiltonian, as explained in \cite{Morley19a} is that one can effectively ``scale to infinity'' all behaviour except for directly at the avoided crossing. For this reason, the boundaries are moved to $f(0)=\infty$ and $f(1)=-\infty$. An advantage of this model is that it applies equally well to the ground and first excited state of any setting where the evolution becomes dominated by a single avoided crossing. This applies to both drivers we have discussed here, and much more widely to many other cases \cite{Chakraborty2016randGraphs}. As an example of the approach to single avoided crossing behaviour, figure \ref{fig:hyp20_ac_illustration} shows the spectrum of the (qubit) hypercube Hamiltonian for $n=20$, a size which already strongly shows the limiting behaviour. A minor disadvantage of the single avoided crossing model is that it loses information about constant factors away from the asymptotic limit (as were examined in detail in \cite{Morley19a}). Since the purpose of our current work is to argue the existence of speedups, not to compare models, this is not a major drawback. 

Since it only describes a two-state Hilbert space, the single avoided crossing model can be diagonalised by hand. The energy gap between the ground and first excited state is 
\begin{equation}
    g(\tau)=g_\mathrm{min}\sqrt{f^2(\tau)+1} \label{eq:gap_ac_app},
\end{equation}
defining $\tau=\frac{t}{T}$ for mathematical convenience.

The ground and first excited states are 
\begin{align}
    \ket{g}=\frac{1}{\mathcal{N}(\tau)}\left(\ket{0}+ \left(\sqrt{f^2(\tau)+1}+f(\tau)\right)\ket{1} \right), \label{eq:ground_state}\\
    \ket{e}=\frac{1}{\mathcal{N}(\tau)}\left(\left(\sqrt{f^2(\tau)+1}+f(\tau)\right)\ket{0}- \ket{1} \right),\label{eq:excited_state}
\end{align}
where $\mathcal{N}(\tau)$ is a normalisation factor,
\begin{equation}
    \mathcal{N}(\tau)=\sqrt{\left(\sqrt{f^2(\tau)+1}+f(\tau)\right)^2+1}. \label{eq:normalisation}
\end{equation}

We apply the adiabatic condition and solve following the method of \cite{Roland2002} (see also \cite{Morley19a}). The differential equation for $f(\tau)$, where we have defined $\tau=\frac{t}{T}$ for mathematical convenience, becomes
\begin{equation}
\left|\frac{\partial f(\tau)}{\partial \tau}\right|\propto\frac{g^2(\tau)}{\left| \left<\frac{\partial H_{\mathrm{ac}}}{\partial f} \right>_{0,1}\right|}=\frac{g_{\mathrm{min}}}{4}\left(1+f^2(\tau)\right),
\end{equation}
where $g(\tau)$ is the energy gap between the ground and first excited state defined in equation \ref{eq:gap_ac_app}. We have mapped the initial state $\ket{\tilde{\omega}}\rightarrow \ket{1}$ in this model and $\ket{m}\rightarrow \ket{0}$. Applying the boundary conditions and solving, we obtain,
\begin{equation}
f(\tau)=\cot(\pi \tau). 
\end{equation}
Recasting in terms of $\tau$ we note an important property: The only dependence on $g_\mathrm{min}$ is a global multiplicative factor, therefore if we consider the instantaneous action of the Hamiltonian at $\tau$ acting for a short time $\delta t$ and setting the total runtime $T= t_\mathrm{scale} g^{-1}_\mathrm{min}$
\begin{equation}
    H_{\mathrm{ac}} \delta t=\delta \tau \frac{T g_\mathrm{min}}{2}\left[f(\tau)Z-X\right]=\delta \tau t_\mathrm{scale}\left[f(\tau)Z-X\right]
\end{equation}
 this scaling removes all dependence of time evolution on the minimum gap. The physical meaning of this is that a longer runtime can perfectly compensate for slower dynamics, since the timescale of the dynamics (defined by the minimum gap) scales with $N^{-\frac{1}{2}}$, the runtime must scale as $N^\frac{1}{2}=\sqrt{N}$ to compensate, a scaling which corresponds to the optimal Grover speedup. This is a somewhat generalised version of the result from \cite{Roland2002}. It is worth observing that in additional to the optimal adiabatic schedule, any schedule which is defined to be scale-invariant and results in a non-zero final overlap with $\ket{m}$ will also exhibit optimal $\sqrt{N}$ scaling. These include continuous-time quantum walk ($f(\tau)=0$) and quantum walk/adiabatic interpolations as defined in \cite{Morley19a}. Interestingly, this adiabatic schedule can be related to the Gauss-Bonnet theorem in the vicinity of the anti-crossing \cite{dridi2019homological}. This suggests a differential geometry description of the different Zeno manifestations presented in this paper. We leave this for future work.

\section{Operating principles of Zeno and non-Zeno algorithms\label{appendix:alg_op}}

\subsection{Purity arguments for measurements and decoherence}
In the case we are considering, in which the density matrices are two-dimensional, the density matrix can always be written as 
\begin{equation}
    \rho=p_\psi \ketbra{\psi}{\psi}+p_m \ketbra{m}{m}
\end{equation}
where $p_\psi+p_m=1$, $p_\psi \ge 0$, $p_m\ge 0$ and $\ket{\psi}$ is a normalised quantum state vector such that $0 < |\braket{\psi}{m}|$. Any pure state $\ketbra{\psi}{\psi}$ can be represented for example by setting $p_\psi=1$, $p_m=0$, and the maximally mixed state can be represented by setting $\ket{\psi}=\ket{\tilde{m}}$ and $p_m=p_\psi=1/2$. To decompose a general density matrix into this form, we observe that a complete set of equations uniquely defining the decomposition take the form, 
\begin{align}
\sandwich{m}{\rho}{m}=p_m+p_\psi|\braket{\psi}{m}|^2,\\
\sandwich{\tilde{\omega}}{\rho}{m}=p_\psi\braket{\psi}{m}\braket{\tilde{\omega}}{\psi},\nonumber \\
\sandwich{\tilde{\omega}}{\rho}{\tilde{\omega}}=p_\psi|\braket{\psi}{\tilde{\omega}}|^2,\\
\braket{\psi}{\psi}=1.
\end{align}
From these equations, it follows that,
\begin{align}
\mathrm{Tr}\left[\rho^2\right]=p^2_\psi\left(\left|\braket{\psi}{\tilde{\omega}}\right|^2+\left|\braket{\psi}{m}\right|^2\right)+p^2_m=\\ p^2_\psi+p^2_m=\left(1-p_m\right)^2+p^2_m=2\,p^2_m-2p_m+1.
\end{align}
Applying the quadratic formula then gives 
\begin{equation}\label{eq:pm}
    p_m=\frac{1}{2}\left(1\pm\sqrt{1-2\left(1-\mathrm{Tr}\left[\rho^2\right]\right)}\right).
\end{equation}
In the pure case in which $\mathrm{Tr}\left[\rho^2\right]=1$, Eqn~\eqref{eq:pm} becomes $p_m=(1\pm 1)/2$, so $p_m$ is just $0$ or $1$. This makes intuitive sense: a pure state could be the state $\ket{m}$ (corresponding to the value of $1$), or any other state (corresponding to the value of $0$), but any non-trivial sum of the two will no longer be pure\footnote{Recall that the probability to be measured in $\ket{m}$ is not simply $p_m$, but $p_m+p_\psi\left|\braket{m}{\psi}\right|^2$ so $p_m$ is necessarily zero for pure states with non-unity probability to be found in the state $\ket{m}$}. On the other extreme, a maximally mixed state, $\mathrm{Tr}\left[\rho^2\right]=1/2$, $p_m=1/2$. For intermediate values, this formula gives a non-zero lower bound on the value of $p_m$, since the negative branch will yield a smaller $p_m$ value for $1/2\le\mathrm{Tr}\left[\rho^2\right]<1$, it follows that $p_m\ge \frac{1}{2}\left(1-\sqrt{1-2\left(1-\mathrm{Tr}\left[\rho^2\right]\right)}\right)$. Finally, this can be used to bound the total probability of measuring the marked state,
\begin{align}
    \sandwich{m}{\rho}{m}=p_m+p_\psi\left|\braket{m}{\psi}\right|^2\ge p_m \ge \frac{1}{2}\left(1-\sqrt{1-2\left(1-\mathrm{Tr}\left[\rho^2\right]\right)}\right). 
\end{align}

\subsection{Ground states preserved by Zeno effects}

As discussed in the main text, the previous argument does not strictly apply to the setting of a Zeno effect, since the Zeno effect describes repeated measurements which maintain a pure state. Since rotation of the underlying state space in an avoided crossing is different from the traditional setting of ``freezing'' time evolution, it is worth reviewing how to derive the fact that rapid removal of off-diagonal density matrix elements in the instantaneous energy eigenbasis can maintain the ground state.

To start with, we consider the overall effect of an avoided crossing if no dynamics are present within the instantaneous energy eigenbasis. Since the avoided crossing corresponds to an effective swap of the role of the ground and first excited state, this operation (which corresponds to applying an identity in a fixed basis) will take the form (in terms of $\tilde{Y}=i\ketbra{e}{g}-i\ketbra{g}{e}$) 
\begin{equation}
    \lim_{q\rightarrow\infty} \mathcal{T}\prod^q_j\left( \mathbb{1}\sqrt{1-\delta^2_j}-i\delta_j \tilde{Y}\right)=\lim_{q\rightarrow \infty}\mathcal{T} \prod^q_j\left( \mathbb{1}-i\delta_j \tilde{Y}\right) \label{eq:basis_Y}
\end{equation}
Where $\delta_j$ is the instantaneous change in the eigenbasis, defined as
\begin{align}
\delta_j=\braket{e\left(\tau=\frac{j}{q}\right)}{g\left(\tau=\frac{j-1}{q}\right)}\\
=\frac{1}{\mathcal{N}\left(\frac{j}{q}\right)\mathcal{N}\left(\frac{j-1}{q}\right)}\nonumber \\ \left[\sqrt{f^2\left(\frac{j}{q}\right)+1}+f\left(\frac{j}{q}\right)-\left(\sqrt{f^2\left(\frac{j-1}{q}\right)+1}+f\left(\frac{j-1}{q}\right) \right)\right] \nonumber \\ \approx \frac{1}{q\mathcal{N}^2\left(\frac{j}{q}\right)}\frac{\partial }{\partial x}\left( \sqrt{f^2(x)+1}+f(x)\right){\huge\big|}_{x=\frac{j}{q}}\\=\frac{1}{q\mathcal{N}^2\left(\frac{j}{q}\right)}\frac{\partial f(x)}{\partial x}{\huge\big|}_{x=\frac{j}{q}}\left(\frac{f\left(\frac{j}{q}\right)}{\sqrt{f^2\left(\frac{j}{q}\right)+1}}+1 \right),
\end{align}
and the equality in equation \ref{eq:basis_Y} follows from the fact that $\delta_j\rightarrow 0$ in the limit, and therefore terms of order $\delta^2_j$ can be neglected. The term $\mathcal{N}$ is the normalisation factor from equation \ref{eq:normalisation}.
For a single $j$ value the action of this operation on an arbitrary (Hermitian so $\rho_{eg}=\rho^\star_{ge}$) density matrix is
\begin{align}
    \left( \mathbb{1}-i\delta_j \tilde{Y}\right)\left(\rho_{gg}\ketbra{g}{g}+\rho_{ge}\ketbra{g}{e}+\rho^\star_{ge}\ketbra{e}{g}+\rho_{ee}\ketbra{e}{e}\right)\left( \mathbb{1}+i\delta_j \tilde{Y}\right)= \label{eq:exp_step}\\
    \left(\rho_{gg}-2 \delta_j\mathrm{Re}[\rho_{ge}]+\delta^2_j\rho_{ee}\right)\ketbra{g}{g}+\left(\rho^\star_{ge}+\delta_j(\rho_{gg}-\rho_{ee})+\delta^2_j\rho_{ge}\right)\ketbra{e}{g}+\nonumber \\
    \left(\rho_{ge}+\delta_j(\rho_{ee}-\rho_{gg})+\delta^2_j\rho^\star_{ge}\right)\ketbra{g}{e}+\left(\rho_{ee}+2 \delta_j\mathrm{Re}[\rho_{ge}]+\delta^2_j\rho_{gg}\right)\ketbra{e}{e}=\nonumber \\
    \left(\rho_{gg}-2 \delta_j\mathrm{Re}[\rho_{ge}]\right)\ketbra{g}{g}+\left(\rho^\star_{ge}+\delta_j(\rho_{gg}-\rho_{ee})\right)\ketbra{e}{g}+\nonumber \\
    \left(\rho_{ge}+\delta_j(\rho_{ee}-\rho_{gg})\right)\ketbra{g}{e}+\left(\rho_{ee}+2 \delta_j\mathrm{Re}[\rho_{ge}]\right)\ketbra{e}{e}+O(\delta^2_j). 
\end{align}
From this formula, we can immediately see that if we start in the ground state $\rho_{gg}=1$ and $\rho_{ee}=\rho_{ge}=0$ and a phase is introduced to $\rho_{ge}$ at each step, than the sign of $\mathrm{Re}[\rho_{ge}]$ will eventually flip causing, the flow of amplitude to reverse, flowing instead from the excited state to the ground state, this is the underlying mechanism behind both adiabatic operation and the multi-stage quantum walk manifestation of the Zeno effect. An extreme manifestation occurs when phase flipping is studied, in which case the flow of probability reverses direction in a single step, returning the lost probability to the ground state.

If we instead consider the operation in equation \ref{eq:exp_step} with a measurement in the energy eigenbasis at each step (sending $\rho_{ge}\rightarrow 0$), we have 
\begin{align}
    \left( \mathbb{1}-i\delta_j \tilde{Y}\right)\left(\rho_{gg}\ketbra{g}{g}+\rho_{ee}\ketbra{e}{e}\right)\left( \mathbb{1}+i\delta_j \tilde{Y}\right)= \label{eq:exp_step_dec}\\
    \left(\rho_{gg}+\delta^2_j\rho_{ee}\right)\ketbra{g}{g}+\delta_j(\rho_{gg}-\rho_{ee})\ketbra{e}{g}+ \nonumber \\
    \delta_j(\rho_{ee}-\rho_{gg})\ketbra{g}{e}+\left(\rho_{ee}+\delta^2_j\rho_{gg}\right)\ketbra{e}{e}=\\
    \mathbb{1}+\delta_j(\rho_{ee}-\rho_{gg})\ketbra{g}{e}+\delta_j(\rho_{gg}-\rho_{ee})\ketbra{e}{g}+O(\delta^2_j),
\end{align}
if we now start in the ground state $\rho_{gg}=1$ and $\rho_{ee}=\rho_{ge}=0$ and perform successive operations where $\rho_{ee}\rightarrow \rho_{ee}+\delta^2_j\rho_{gg}$, we find that $\rho_{ee}=\sum_j\delta^2_j\approx 0$. This illustrates the presence of a Zeno effect, which maintains the instantaneous ground state since $\rho_{ee}+\rho_{gg}=1$ by definition. We further note that removal of the excited state at each step does not fundamentally change this picture, it just adds an additional step where $\ket{e}\rightarrow \ket{d}$, and therefore similarly $\rho_{dd}=\sum_j\delta^2_j\approx 0$. These arguments apply equally to rapid continuous dissipation or decoherence.

\section{Derivation of master equations \label{appendix:mast_derive}}

\subsection{Derivation of equation \ref{eq:dephase_lindblad}\label{appendix:eq:dephase_lindblad}}

The first stage of the derivation is to consider the average case of an instance of equation \ref{eq:dephase_evolve} between two $\tilde{X}$ operations. We will assume that this is operation is occurring after many flips have occurred, so there is a $50$\% chance that an odd number of $\tilde{X}$ operations have been applied and a $50$\% chance an even number have been applied. During this time period the evolution will be described as

\begin{align}
U_\mp(q,j_0)=\mathcal{T}\prod^{j_0+ m/q}_{j=j_0}\exp\left(\frac{\mp i t_{\mathrm{scale}} \sqrt{\gamma^2_j+1}}{m}\tilde{Z}\right)=\nonumber \\ \mathcal{T}\prod^{j_0+ m/q}_{j=j_0}\exp\left(\frac{\mp i t_{\mathrm{scale}} \sqrt{f^2(\tau_j)+1}}{m}\tilde{Z}\right)
\end{align}

where $j_0$ is the starting index, which depends on the point in the evolution and whether the phase is positive or negative depends on the parity of the number of previous applications of $\tilde{X}$, with a negative sign corresponding to an even number, and positive to an odd number. We can then make an approximation based on the fact that $q \gg 1$, we assume that the change in $f(\tau_j)$ is not large over the range of $j$, therefore $f(\tau_j)\approx f(\tau)\quad \forall j$. The expression becomes 

\begin{align}
U_\mp(q,\tau)\approx\mathcal{T}\prod^{j_0+m/q }_{j=j_0}\exp\left(\frac{\mp i t_{\mathrm{scale}} \sqrt{f^2(\tau)+1}}{m}\tilde{Z}\right)= \nonumber \\
\left[\exp\left(\frac{\mp i t_{\mathrm{scale}} \sqrt{f^2(\tau)+1}}{m}\tilde{Z}\right)\right]^{m/q}=\exp\left(\frac{\mp i t_{\mathrm{scale}} \sqrt{f(^2\tau)+1}}{q}\tilde{Z}\right).
\end{align}

For time evolution of a density matrix we should consider the positive and negative cases separately, thus the overall action over a single time period is:

\begin{align}
U_\mp(q,\tau)\rho U^\dagger_\mp(q,\tau)=\nonumber \\  \cos^2(\Phi) \rho\mp  i\cos(\Phi)\sin(\Phi) \tilde{Z} \rho \pm   i\cos(\Phi)\sin(\Phi) \rho \tilde{Z}+ \sin^2(\Phi) \tilde{Z}\rho\tilde{Z}
\end{align}

where

\begin{equation}
    \Phi=-\frac{t_{\mathrm{scale}} \sqrt{f^2(\tau)+1}}{q}.
\end{equation}

Considering a $50$\% chance of evolving with a positive or negative phase, the overall evolution is

\begin{align}
\rho \rightarrow \frac{1}{2}\left(U_-(q,\tau)\rho U^\dagger_-(q,\tau)+U_+(q,\tau)\rho U^\dagger_+(q,\tau)\right)\\
=\rho \cos^2\left(\Phi\right)+\tilde{Z}\rho \tilde{Z}\sin^2\left(\Phi\right).
\end{align}

To derive a master equation, we take the limit where $\Phi$ is relatively small, in other words taking a Taylor expansion which only keeps the first non-trivial terms

\begin{equation}
\rho \rightarrow \rho+\left(\tilde{Z}\rho \tilde{Z}- \rho\right) \Phi^2/2+ O(\Phi^4).
\end{equation}

To convert into a differential equation, we need to also consider the amount of time it has taken for this process to occur which is $t_\mathrm{scale}/q$, this gives

\begin{equation}
    \frac{\Delta \rho}{\Delta t}\approx\frac{\Phi^2q}{2t_\mathrm{scale}}\left(\tilde{Z}\rho \tilde{Z}- \rho\right)=t_\mathrm{scale}\frac{f^2(\tau)+1}{2q}\left(\tilde{Z}\rho \tilde{Z}- \rho\right)=\kappa\left(\tilde{Z}\rho \tilde{Z}- \rho\right),
\end{equation}

matching the more informal arguments in the main text, taking an infinitesimal limit gives 

\begin{equation}
\frac{\partial \rho}{\partial t}=\kappa\left(\tilde{Z}\rho \tilde{Z}- \rho\right)
\end{equation}

which is equation \ref{eq:dephase_lindblad}. An astute reader may be concerned that this derivation assumes that many $\tilde{X}$ operations have already occurred. Recall however that at the initial points of the evolution the ground and excited states will be unchanged and to a very good approximation $\ket{\tilde{\omega}}$ and $\ket{m}$, with the system starting in $\ket{\tilde{\omega}}$. In this case, the relative phase between the ground and first excited state is fairly irrelevant, allowing time for randomisation in the phase directions to build up before the approach to the crossing.

\subsection{Derivation of equation \ref{eq:dissipate_lindblad} \label{appendix:eq:dissipate_lindblad}}

As with the previous derivation, we start by arguing that in the limit of large $q$ we can consider the Hamiltonian essentially static, in other words we can approximate 

\begin{align}
U(t_\mathrm{rot},t_\mathrm{scale})=\mathcal{T}\prod_{j=1}^{m}U^\dagger_\mathrm{meas}(\frac{t_\mathrm{scale}}{mq},\tau_{k,j})U_\mathrm{rot}(t_\mathrm{rot})U_\mathrm{meas}(\frac{t_\mathrm{scale}}{mq},\tau_{k,j}) \approx \nonumber \\
\left[ U^\dagger_\mathrm{meas}(\frac{t_\mathrm{scale}}{mq},\tau_{k,0})U_\mathrm{rot}(t_\mathrm{rot})U_\mathrm{meas}(\frac{t_\mathrm{scale}}{mq},\tau_{k,0})\right]^m= \nonumber \\
 U^\dagger_\mathrm{meas}(\frac{t_\mathrm{scale}}{mq},\tau_{k,0})U^m_\mathrm{rot}(t_\mathrm{rot})U_\mathrm{meas}(\frac{t_\mathrm{scale}}{mq},\tau_{k,0}) = \nonumber \\
 \ketbra{g}{g}\otimes \mathbb{1} \otimes \mathbb{1}_2+\ketbra{e}{e}\otimes \mathbb{1} \otimes \exp(-i 2  t_\mathrm{rot} t_\mathrm{scale}\frac{\sqrt{f^2(\tau)+1}}{q} X),
\end{align}

we now consider the additional action of the measurement, post-selected to move to the ``destroyed'' state $\ket{d}$, which is considered a failure of the algorithm.

\begin{equation}
\Pi_{d1}=\mathbb{1}_2\otimes \mathbb{1}\otimes\ketbra{d}{1}+\mathbb{1}_2\otimes \mathbb{1} \otimes \ketbra{0}{0}
\end{equation}

The overall action of evolution plus measurement plus post selection is (neglecting the continuous system where the overall action is the identity)

\begin{align}
\Pi_{d1}U(t_\mathrm{rot},t_\mathrm{scale}) \nonumber \\ \nonumber= \Pi_{d1} \left[\ketbra{g}{g} \otimes \mathbb{1}_2+\ketbra{e}{e} \otimes \exp(-i 2  t_\mathrm{rot} t_\mathrm{scale}\frac{\sqrt{f^2(\tau)+1}}{q} X) \right]= \nonumber \\
\ketbra{g}{g}\otimes \mathbb{1}_2+\ketbra{e}{e} \otimes \left(\cos\left(\Phi\right) \ketbra{0}{0}-i\sin\left(\Phi\right) \ketbra{d}{0}\right)= \nonumber \\
\ketbra{g}{g}\otimes \mathbb{1}_2+\ketbra{e}{e}\otimes\left(\left(\mathbb{1}-\frac{1}{2}\Phi^2+ O(\Phi^4)\right)\ketbra{0}{0}-i\left(\Phi+O(\Phi^3) \right)\ketbra{1}{1}\right)
\end{align}

where on the last line we have omitted terms where the auxilliary system starts in the $\ket{1}$ state, since it will always be initialised as $\ket{0}$. We have defined
\begin{equation}
   \Phi= 2  t_\mathrm{rot} t_\mathrm{scale}\frac{\sqrt{f^2(\tau)+1}}{q}.
\end{equation}
for notational convenience. We now consider the effect of taking a projective measurement to determine if the auxilliary system is in the $\ket{d}$ state, and abort the algorithm if it is. For notational simplicity, we consider an aborted algorithm as placing the entire system (as opposed to just the auxilliary) into a computationally useless $\ket{d}$ state. By the Born measurement rule, the probability of aborting the algorithm is $\sin^2\left(\Phi \right)\approx \Phi^2$. Since the system can only be put into the $\ket{d}$ state if it started in the $\ket{e}$ state, the overall effect on the density matrix is,
\begin{equation}
    \rho\rightarrow \rho+\left(\ketbra{d}{e}\rho \ketbra{e}{d}-\frac{1}{2}\ketbra{e}{e}\rho-\frac{1}{2}\rho\ketbra{e}{e}\right)\Phi^2+O(\Phi^4).
\end{equation}

As before to convert to a master equation, we consider 
\begin{align}
    \frac{\Delta \rho}{\Delta t}\approx\left(\ketbra{d}{e}\rho \ketbra{e}{d}-\frac{1}{2}\ketbra{e}{e}\rho-\frac{1}{2}\rho\ketbra{e}{e}\right)\frac{\Phi^2 }{\frac{t_\mathrm{scale}}{q}+t_\mathrm{rot}}\approx \nonumber \\ \left(\ketbra{d}{e}\rho \ketbra{e}{d}-\frac{1}{2}\ketbra{e}{e}\rho -\frac{1}{2}\rho\ketbra{e}{e}\right)\frac{\Phi^2 q}{t_\mathrm{scale}}= \nonumber \\
    \frac{4t^2_\mathrm{rot}t_\mathrm{scale}(f^2(\tau)+1)}{q}\left(\ketbra{d}{e}\rho \ketbra{e}{d}-\frac{1}{2}\ketbra{e}{e}\rho-\frac{1}{2}\rho\ketbra{e}{e}\right)= \nonumber \\
    \kappa \left(\ketbra{d}{e}\rho \ketbra{e}{d}-\frac{1}{2}\ketbra{e}{e}\rho-\frac{1}{2}\rho\ketbra{e}{e}\right),
\end{align}
taking a limit of small $\delta t$ gives a result which matches the more informal derivation of the main text,
\begin{equation}
    \frac{\partial \rho}{\partial t}=\kappa \left(\ketbra{d}{e}\rho \ketbra{e}{d}-\frac{1}{2}\left(\ketbra{e}{e}\rho+\rho\ketbra{e}{e}\right)\right).
\end{equation}

\section{Equation \ref{eq:disp_op} in the density matrix formalism and an alternative model for decoherence \label{appendix:dec_model}}

Conceptually it is clear that decoherence operations could be performed by replacing the projection operation in equation \ref{eq:disp_op} with an operation which removes the phase coherence between the two states and resets the $\ket{1}$ state on the additional qubit back to $\ket{0}$ in the case of a $1$ result of the measurement. 

Mathematically, however, this requires conversion of equation \ref{eq:disp_op} to the density matrix formalism, since we cannot represent a system subject to decoherence as an un-normalised state vector. The initial state can be represented as
\begin{equation}
\rho_{\mathrm{start}}=\rho_\mathrm{sys}\otimes \left(\ketbra{x=0}{x=0}\right)\otimes \ketbra{0}{0} \label{eq:rho_init}
\end{equation}
where $\rho_\mathrm{sys}$ is the initial state of the avoided crossing system. By applying this equation to both sides, and including the destroyed state explicitly, equation \ref{eq:disp_op} becomes
\begin{equation}
\rho_\mathrm{fin}=U^\dagger_\mathrm{meas}(t_\mathrm{scale},\tau)\Pi_{d1} U_\mathrm{rot}(t_\mathrm{rot})U_\mathrm{meas}\rho_{\mathrm{start}}U^\dagger_\mathrm{meas}U^\dagger_\mathrm{rot}(t_\mathrm{rot})\Pi^\dagger_{d1}U_\mathrm{meas}(t_\mathrm{scale},\tau), \label{eq:disp_op_dm}
\end{equation}
where 
\begin{equation}
\Pi_{d1}=\mathbb{1}_2\otimes \mathbb{1}\otimes\ketbra{d}{1}+\mathbb{1}_2\otimes \mathbb{1} \otimes \ketbra{0}{0}
\end{equation}
is the mathematical operation which sends the system to the $\ket{d}$ state if the additional qubit is in the $\ket{1}$ state and leaves it unchanged otherwise. The remaining degrees of freedom are left for mathematical convenience, for our purposes any system where the last degree of freedom is in the $\ket{d}$ state is considered computationally useless\footnote{To avoid this construction one can instead construct a superoperator which acts on $\rho$ and separately projects each possible state of the system into a global $\ketbra{d}{d}$ state, but we have elected to maintain these extra degrees of freedom to simplify the expression.}. Any operation is considered to act as the identity on the $\ket{d}$ state, in other words $U\rightarrow U\oplus \mathbb{1}_1$ for all of the unitary operations within equation \ref{eq:disp_op_dm} where the one-dimensional identity acts on the destroyed state which has been added to the Hilbert space.

We note that, aside from the amplitude sent to the $\ket{d}$ state, $\rho_\mathrm{fin}$ from equation \ref{eq:disp_op_dm} has the same structure as equation \ref{eq:rho_init}, in other words the continuous variable and the additional qubit have been reset to the same value.

If we now consider the case without the $\ket{d}$ state, we can also represent decoherence (time dependence dropped to fit on a single line) 
\begin{equation}
\rho_\mathrm{fin}=U^\dagger_\mathrm{meas}\left(\Pi_0+X_\mathrm{add}\Pi_1 \right) U_\mathrm{rot}U_\mathrm{meas}\rho_{\mathrm{start}}U^\dagger_\mathrm{meas}U^\dagger_\mathrm{rot}\left(\Pi_0+\Pi_1 X_\mathrm{add} \right)U_\mathrm{meas}.
\label{eq:decohere_alt}
\end{equation}
In this equation we have defined $X_\mathrm{add}=\mathbb{1}_2\otimes \mathbb{1}\otimes X$ and as usual $\Pi_0=\mathbb{1}_2\otimes \mathbb{1}\otimes \ketbra{0}{0}$, $\Pi_1=\mathbb{1}_2\otimes \mathbb{1}\otimes \ketbra{1}{1}$. In this case, $\rho_\mathrm{fin}$, takes the form of equation \ref{eq:rho_init}, and the effect of the operation is to reduce the off-diagonal elements of $\rho_\mathrm{sys}$.

\bibliography{reference}  

\end{document}